\documentclass[twocolumn]{aastex6}	% 
\usepackage{color}

\slugcomment{To appear in MNRAS}
\shorttitle{Proxima Cen Stellar Cycle}
\shortauthors{Wargelin et al.}

\newcommand{\chandra}{{\it Chandra}}
\newcommand{\xmm}{{\it XMM-Newton}}
\newcommand{\rosat}{{\it ROSAT\,}}

\newcommand{\asca}{{\it ASCA}}

\newcommand{\exosat}{{\it EXOSAT}}
\newcommand{\einstein}{{\it Einstein}}

\newcommand{\swift}{{\it Swift}}
\newcommand{\iue}{{\it IUE}}
\newcommand{\fuse}{{\it FUSE}}

\newcommand{\caii}{Ca\,{\sc ii}}
\newcommand{\mgii}{Mg\,{\sc ii}}
\newcommand{\nai}{Na\,{\sc i}}
\newcommand{\hei}{He\,{\sc i}}

\begin{document}
%\title{A \swift\ Long Look at Proxima Centauri:\\
%Flares, Spectra, and Cycles}
\title{Optical, UV, and X-Ray Evidence for a 7-Year\\
Stellar Cycle in Proxima Centauri}

\author{B.~J.\ Wargelin\altaffilmark{1},
S.~H.\ Saar\altaffilmark{1}, 
G.~Pojma{\'n}ski\altaffilmark{2},
J.~J.\ Drake\altaffilmark{1},
and V.~L.\ Kashyap\altaffilmark{1}}
\altaffiltext{1}{
%Smithsonian Astrophysical Observatory, 
Harvard-Smithsonian Center for Astrophysics, 
60 Garden Street, MS-70, 
Cambridge, MA 02138, USA; bwargelin@cfa.harvard.edu}
\altaffiltext{2}{Astronomical Observatory University of Warsaw,
Al.~Ujazdowskie 4, 00-478, Warszawa, Poland}

%%%%%%%%%%%%%%%%%%%%%%%%%%%%%%%%%%%%%%%%%%%%%%%%%%%%%%%%%%%%%%%%%%%%

\begin{abstract}
% 250 word count limit
Stars of stellar type later than about M3.5 are believed to be fully
convective and therefore unable to support magnetic dynamos like
the one that produces the 11-year solar cycle.  Because of their intrinsic
faintness, very few late M stars have undergone long-term monitoring
to test this prediction, which is critical to our understanding
of magnetic field generation in such stars.
Magnetic activity is also of interest as the driver of
UV and X ray radiation,
as well as energetic particles and stellar winds,
that affect the atmospheres
of close-in planets that lie within habitable zones,
such as the recently discovered Proxima b.
We report here on several years of
optical, UV, and X-ray observations of Proxima Centauri (GJ 551; dM5.5e):
15 years of ASAS photometry in the
$V$ band (1085 nights)
and 3 years in the $I$ band (196 nights),
4 years of \swift\ XRT and UVOT observations (more than 120 exposures),
and 9 sets of X-ray observations from other X-ray missions
(\asca, \xmm, and three \chandra\ instruments)
spanning 22 years.
We confirm previous reports of an 83-day rotational period
and find strong evidence for a 7-year stellar cycle,
along with indications of differential rotation at about the solar level.
X-ray/UV intensity is anti-correlated with optical $V$-band brightness
for both rotational and cyclical variations.
From comparison with other stars observed to have X-ray cycles
we deduce a simple empirical relationship between X-ray cyclic modulation
and Rossby number,
and we also present
\swift\ UV grism spectra covering 2300--6000 \AA.
\end{abstract}

% In alphabetical order--see online list in ApJ For Authors/Suject key words
% http://authors.iop.org.ezp-prod1.hul.harvard.edu/atom/usermgmt.nsf/AuthorServices?OpenForm&ISSN=0004-637X

%%%%%%%%%%%%%%%%%%%%%%%%%%%%%%%%%%%%%%%%%%%%%%%%%%%%%%%%%%%%%
%% sections all caps
%% subsection, subsubsection, paragraph--caps for 1st letter each word except
%% articles and prepositions

%% /citep for parenthetical citation
%% /citet to make citation part of the sentence
%% /citep{cit:matthews1993,cit:anosova1994}, 

%%%%%%%%%%%%%%%%%%%%%%%%%%%%%%%%%%%%%%%%%%%%%%%%%%%%%%%%%%%%%%%%%%%%%%
\section{INTRODUCTION}
\label{sec:intro}
%%%%%%%%%%%%%%%%%%%%%%%%%%%%%%%%%%%%%%%%%%%%%%%%%%%%%%%%%%%%%%%%%%%%%%

%P1
Stellar activity cycles, seen in the Sun and many late-type stars,
are driven by magnetic activity and therefore reflect
a star's magnetic field strength, internal structure, rotation, and evolution.
Studying those cycles
%into the generation and dynamics of stellar magnetic fields.  
can provide key information on the dynamo process, which powers
magnetic regeneration in stars, accretion disks, and planets.
Many details of that process, however, remain poorly understood,
even for the 11-yr solar cycle.
%Despite much recent progress,
%even the 11-yr solar cycle remains incompletely understood.
Observational comparisons with other stars are therefore vital
for constraining models of magnetic activity and explaining the
presence or lack of stellar cycles.
%and data on stellar cycles in different environments helps 
%to constrain theories of their magnetic origins.

%P2
Understanding stellar magnetic activity is also relevant
to studies of exoplanets because starspots and flares
can mimic or obscure the signatures of planets 
\citep{cit:queloz2001}
and may affect those planets' habitability.
% If a ref is needed for above use Barnes 2001 MNRAS.
This latter subject is especially interesting 
in light of the recent discovery of
an exoplanet orbiting 
in the habitable zone of our Sun's nearest neighbor, 
Proxima Centauri. Proxima b has a minimum mass of about 1.3 times that of the 
Earth and an orbital period of 11.2 days with a semi-major axis of only 
0.049~AU, about one-eighth Mercury's orbital radius 
\citep{cit:Anglada2016}.
A key factor in planetary habitability is the effect on the atmosphere
of X-ray/UV radiation and the stellar wind
\citep[e.g.,][and references therein]{cit:lammer2003,
cit:khodachenko2007,
cit:penz2008,
cit:cohenetal2015,
cit:owen2016}
which are ultimately driven by the stellar magnetic field.

Cycles in most cool stars (F--M) are thought to arise
from the interplay of
large scale shear (differential rotation) and small scale helicity in an
$\alpha\Omega$ dynamo.
The current paradigm has the $\Omega$ effect sited in the tachocline
layer at the bottom of the convection zone \citep{cit:dikpati1999}.
Stars later than type $\sim$M3.5
are expected to be fully convective \citep{cit:chabrier1997}
with their magnetic activity probably arising
from the $\alpha^{2}$ process, which is not considered
conducive to generating cyclic behavior,
although some modelers suggest that cycles may be possible
in certain parameter regimes
\citep[e.g.,][]{cit:rudiger2003,cit:gastine2012,cit:kaplyla2013}.
%(e.g., \citet{cit:rudiger2003};
%\citet{cit:gastine2012};
%\citet{cit:kaplyla2013}--**Some refs repeated later).
%(e.g., R\"udiger et al.\ 2003; Gastine et al.\ 2012;
%K\"apyl\"a et al.\ 2013).
Evidence has also recently emerged that fully convective stars share 
the same rotation-activity relation as stars with radiative cores, 
supporting the idea that the tachocline is not a key ingredient of 
solar-type dynamos \citep{cit:wright2016}.
Proof of the existence of cycles in fully convective stars
and how their character varies with stellar properties
would greatly advance our understanding of stellar dynamos
but such stars' intrinsic faintness,
often coupled with short term variability that can
mask longer term trends, makes this difficult.

%P4
The longest running program to look for signs of cyclic magnetic
activity is the
HK Project at Mount Wilson Observatory (MWO), which began c.~1966
\citep{cit:wilson1978,cit:baliunas1995} and
monitors chromospheric
\caii\ H and K lines (3969 and 3934 \AA)
as indicators of the strength and covering fraction of
stellar magnetic fields.
This project currently includes about 300 stars of spectral
type F--K, but only a single M star
(HD 95735; Lalande 21185; dM2)
because of the general faintness of M dwarfs
and the relative weakness of their \caii\ lines.
%Despite the difficulty in monitoring Lalande 21185, however,
%using data from 1968 to 2002,
%\citet{cit:baliunas2006} found weak evidence for a $\sim$2.28-yr cycle.

%P5
Other monitoring projects, including HARPS \citep{cit:mayor2003},
%(Mayor et al.~2003),
the McDonald Observatory (MDO) M Dwarf Planet Search 
\citep{cit:cochran1993,cit:endl2003},
%(Cochran \& Hatzes 1993),
the CASLEO/HK$\alpha$ Project \citep{cit:cincu2004},
%(Cincunegui \& Mauas 2004),
the REsearch Consortium On Nearby Stars
\citep[RECONS;][]{cit:henry1994,cit:hosey2015},
%(RECONS; Henry et al.~1994; Hosey et al.~2015),
%{cit:henry1994,cit:hosey2015}
%(RECONS; Winters et al.~2011, 2015),
and the All Sky Automated Survey 
\citep[ASAS;][]{cit:pojmanski1997,cit:pojmanski2002}
%(ASAS; Pojma\'{n}ski 1997, 2002)
%{cit:pojmanski1997 and 2002}
have increased the number of M dwarfs under study,
employing a variety of stellar activity metrics.
These newer programs have now been running for over a decade and
several papers on their early results for M stars
have been published in the past few years.

%P6
Using HARPS spectral data
collected over periods as long as 7 years, 
%Gomes da Silva et al.~(2011, 2012)
\citet{cit:gomes2011,cit:gomes2012}
studied 28 M0--3.5 stars along with Barnard's Star (M4) and
Proxima Cen (M5.5).  
Roughly one third of the stars,
but not Barnard's Star or Prox Cen, showed long
term variability in at least two of the optical lines
studied (\caii, H$\alpha$, He {\sc i} D3, \nai\ D).
%but none later than M3.5.
(Note that the Prox Cen observations, collected during
roughly 40 nights over 6 years, had the lowest signal to noise
ratio of the $\sim$30 stars studied.)
\citet{cit:Anglada2016} used those and newer HARPS data,
along with spectral and photometric data from other instruments,
in their study of Prox Cen and also did not see a cycle,
although they did note roughly 80-d rotational periodicity.

%P7
\citet{cit:robertson2013} 
analysed H$\alpha$ intensities in $\sim$90
M0-M5 stars specifically chosen for their inactivity 
(indicated by a lack of \rosat\ soft X-ray detections) in the MDO program, 
including a dozen M4's and a few M5's.
At least seven stars showed periodicity,
the latest types being M4 for GJ 476 (but listed as type M2.5 in SIMBAD)
and M5 for GJ 581, which \citet{cit:gomes2012} also found to be periodic
but listed as type M2.5.
%GJ 3801, an M4 star, also showed a significant quadratic trend but did not 
%have sufficient temporal coverage to infer a cycle.
GJ 581 was also studied, 
along with 263 other M2--8 stars in the RECONS program, using $VRI$ photometry
by \citet{cit:hosey2015} who did {\it not} see a cycle.
They did, however, find four other stars with
multi-year periodic behavior indicative of a cycle, 
but three of those systems were binaries
and the other cycle was only tentative.
\citet{cit:vida2013}
%Vida, Kriskovics \& Ol\'{a}h (2013)
studied four systems (one K3, one M4, and the others $\sim$M1) 
with very short rotation periods ($\sim$0.45 days) and found 
similarly short cycles ranging from 0.84 to 1.45 yr in all except the M4.
Other nearly-fully convective stars showing signs of a cycle include
AD Leo \citep[M3;][]{cit:buccino2014},
GJ375 \citep[M3;][]{cit:diaz2007},
and perhaps EV Lac \citep[M3.5;][]{cit:alekseev2005}.

%P8
The paucity of results for late-type M stars is not due to
lack of interest, but because of these stars' faintness and
the difficulty of finding suitable activity metrics.
Of the handful of stars with stellar type M4 or later noted above,
the aptly named Proxima Cen 
\citep[dM5.5;][]{cit:bessell1991}
is by far the closest 
\citep[1.305 pc;][]{cit:lurie2014}
and easiest to observe
and several authors have reported indications that it
may have a cycle.
%Benedict et al.\ (1998)
\citet{cit:benedict1998}, analyzing 5 years of photometry data
from the {\it Hubble} Fine Guidance Sensors,
suggested a 3.0-year cycle, though with low confidence.
\citet{cit:cincu2007}, measuring the H$\alpha$ line-to-continuum
on 24 nights over 7 years and excluding obvious flares,
%that showed a $\sim$130\% variation in $S$ (the
%HK Project's line/continuum activity metric)
made Lomb-Scargle periodograms 
%(see Section~\ref{sec:asas})
and found a 1.2-year period with peak-to-peak amplitude
variations of 25\% but a False Alarm Probability of 35\%.
Lastly, \citet{cit:endl2008} found an 
`intriguing peak' in 76 nights
of radial velocity measurements, but the period of that peak 
roughly matches the seven year span (2000--2007) of their observing program,
and they did not see evidence for an
83-d rotation period (see below).
%From their abstract:
%Based on our simulations, we exclude the presence of any planet 
%in a circular orbit with m sin i ≥ 1~M_Neptune at separations of a ≤ 1 AU.

%P9
The most compelling optical evidence for a stellar
cycle in Prox Cen comes from the ASAS project \citep{cit:pojmanski2002},
which monitors millions of stars 
brighter than $\sim$14th magnitude
in the $V$ and $I$ bands
%with $V$-band magnitudes between 7 or 8 and 14.25
% $I$ range is 6.5 to 13.25 mag.
in the southern (beginning 1997)
%(Dec$\la+20^{\circ}$) 
and northern (since 2006) skies.
Currently, $V$-band data from the third of four data collection phases
(ASAS-3; 2000-2010) are available online, along with I-band data
from ASAS-2 (1998-2000, not including Prox Cen).
%($V\sim$11.2 for Prox Cen.)
Using five years of $V$-band data from ASAS-3 supplemented with
UV data from the \iue\ and \fuse\ missions,
%(**Steve gave refs--needed?)
%Jason et al.\ (2007) 
\citet{cit:jason2007} saw indications
of a `probable' cycle of 6.9$\pm$0.5 years in Prox Cen,
later revised to 7.6 years in a \chandra\ observing proposal by
%Guinan (2010)
\citet{cit:guinan2010}.
\citet{cit:savanov2012} later calculated amplitude
power spectra using nine years of ASAS data
and also saw a broad peak around 8 yr,
along with several other peaks at shorter periods.
%(Note later that the Guinan abstract says
%"Also X-ray data from ROSAT, XMM and Chandra show a corresponding 
% coronal X-ray cycle with an expected minimum during 2010/11" because 
% that's just the opposite of what we claim.)
%The latest phase of the ASAS program, ASAS-4, began in 2010
%but those data have not yet been publicly released.
(We learned shortly before acceptance of this paper
that Su\'{a}rez Mascare\~{n}o et al.~2016 (in press)
also analyzed ASAS-3 data and found cycles in seven and perhaps
as many as nine stars of type M4 or later,
including Prox Cen with $P_{\rm rot}=6.8\pm0.3$ yr.)
An activity cycle in a fully convective M star like Prox Cen
would be exciting if confirmed, 
as it would provide evidence that:
%The presence of a cycle in Prox Cen would therefore be evidence
%that either: 
(1) another type of $\alpha\Omega$ dynamo must exist, such as one
driven by shear within the convective zone in the absence
of a tachocline, as suggested in recent models by
\citet{cit:brown2010,cit:brown2011};
%(perhaps one where shear within the convection zone is sufficient); 
(2) there is a magnetically stabilized layer deep in cool M dwarfs
that can act like a tachocline for flux storage/amplification 
\citep[e.g.,][]{cit:mullan2001};
or (3) $\alpha^2$ dynamos can indeed support cycles,
as suggested by some work including
\citet{cit:rudiger2003} and \citet{cit:chabrier2006}.
%e.g., R\"udiger et al.\ 2003; Chabrier \& K\"uker 2006).

%P11
Whether or not Prox Cen has a stellar cycle,
a vital parameter in understanding its magnetic activity
is its rotation rate.
%Several periods have been proposed over the years.
%% Doyle seems to use a rotation-activity curve to predict
%% Prox Cen's rotation rate--not a measurement so I will not mention.
%\citet{cit:doyle1987}
%%, with little detail, 
%suggested a
%period of $51\pm12$ d based on \mgii\ intensities measured by \iue.
\cite{cit:guinan1996} used
\iue\ \mgii\ intensities ($\sim$2800 \AA) from twice-weekly observations over
$\sim$4 months in 1995
to deduce a rotation period of $31.5\pm1.5$ d with 20--25\% variations, 
later revised to $30.5\pm1.5$ \citep{cit:jay1997}, both reported
in conference presentations.
The previously mentioned work by \citet{cit:benedict1998}
using {\it Hubble} FGS data derived a rotation period of 83.5 d 
with 6.6\% (0.069 magnitude) peak-to-peak amplitude
consistent with rotational modulation caused by a single large starspot.
(Smaller variations at half that period were sometimes seen and
ascribed to two starspots $\sim$180$^{\circ}$ apart.
Earlier work by \citet{cit:benedict1993} using a shorter span
of FGS data also found a period of 42 d.)
More recently, \citet{cit:kiraga2007}, \citet{cit:savanov2012},
and \citet{cit:suarez2016}, using between 5 and 9 years of ASAS data,
all derived periods of 83 d.
%More recently, \citet{cit:kiraga2007} derived very similar results
%from a more extensive sample of five years of ASAS data, 
%measuring a period of 82.5 d, which was further confirmed by
%\citet{cit:savanov2012} using four additional years of data.
\citet{cit:reiners2008} comment that this period is longer than
expected given Prox Cen's activity level and
the magnetic field strength of $\sim$600 G that they inferred
from Zeeman broadening in high-resolution spectra.

%%%%%%%%%%%%%%%%%%%%%%%%%%%%%%%%%%%%%%%%%%%%%%%%%%%%%%%%%%%%%%%%%%%%%%
\subsection{X-Ray Period Monitoring}
\label{sec:intro-xrays}

As noted above, M stars are in general quite faint in the optical band,
and any variations in Prox Cen's optical emission caused by
magnetic activity cycles are likely to be at the few percent level.
UV/X-ray emission, however, is a much more sensitive indicator
%than \caii\ line/continuum ratios 
of magnetic activity in late-type stars, particularly M dwarfs.
In the Geostationary Operational Environmental Satellite (GOES)
1--8 \AA\ (1.5--12 keV) band used to monitor solar emission, 
the quiescent X-ray flux ($L_{\rm X}$) varies by two or three orders
of magnitude over a cycle
\citep{cit:wagner1988,cit:aschwanden1994}, 
depending on how stringently
flares are filtered out.
\citet{cit:judge2003}
estimate that in the softer \rosat\ band (0.1--2 keV) 
the Sun's $L_{\rm X}$ varies by a factor of $\sim$6 over a cycle,
while optical-band amplitudes are of order 0.1\%.
% Judge also say "A factor of 1.5 peak-to-peak variation in the RASS 
% passband is predicted due simply to rotational modulations."

A major challenge with X-ray monitoring, however, is
maintaining a sustained campaign by a single mission.
Long gaps in temporal coverage hamper periodicity analyses,
%particularly with respect to phasing (**true?  phrasing OK?),
and instrumental responses can differ a great deal from one
telescope to another, making comparisons problematic.
An example is provided by $\alpha$ Cen (G2V$+$K1V).
%which has been been observed often enough in X-rays to
%be considered `monitored.'
%(Robrade, Schmitt, and Favata 2005).
\citet{cit:ayres2008} reported that an apparent
rapid decrease in X-ray emission from
$\alpha$ Cen A was greatly exaggerated
by energy dependent differences among the multiple
satellite/instrument configurations used to observe it.
\citet{cit:ayres2009} followed that paper with a painstaking analysis that
combined 13 years of X-ray data from three missions and five instruments,
and derived a tentative cycle in $\alpha$ Cen B of nine years,
in agreement with an estimate of 8.36 years derived from
\mgii\ and \caii\ emission by 
%Buccino and Mauas (2008).
\citet{cit:buccino2008}
and an 8.8-yr period found by \citet{cit:dewarf2010}.
%and with an 8.84-year period derived from IUE spectra by
%DeWarf, Datin, and Guinan (2010).
A follow-on paper 
\citep{cit:ayres2014}
that included nine more twice-yearly \chandra\ HRC-I
observations refined the X-ray period to $8.1\pm0.2$ years
with a factor of 4.5 intensity variation,
% Ayres2014  pg 10, R col, for 0.2-2 keV
% Also says alpha Cen A contrast is roughly a factor of 3.
% His Fig 6 plots solar 0.2-2 keV, with Max/min between
% 3.5 and 7.5.  Judge's estimate of about a factor of 6 sounds OK.
and also suggested that $\alpha$ Cen A may have
a $\sim$19-yr cycle.

%Similar X-ray periodicity has been
%seen in three other late-type stars, all of them with stellar cycles
%previously known from optical monitoring:
%61 Cygni A (K5V), Hempelmann et al.\ (2006);
%HD 81809 (G2+G9), Favata et al.\ (2008);
%and $\alpha$Cen (G2V+K1V), Ayres et al.\ (2008).

Three other late-type stars have also been reported to have X-ray cycles.
\citet{cit:hempelmann2006}
studied X-ray data on 61 Cygni A (K5V) from
\rosat\ (8 measurements over 1993--1997) and
\xmm\ (8 over 2002--2005)
and observed X-ray fluxes vary by more than a factor of two over
a 7.3-year cycle,
in agreement with 40 years of \caii\ measurements.
The latest update
\citep{cit:robrade2012} 
reports the same cycle period with factor-of-three intensity variations
over ten years of twice-yearly \xmm\ observations.
(\xmm\ is also monitoring $\alpha$ Cen but the A and B
components are not well resolved.)

The next X-ray cycle measurement is by
\citet{cit:favata2008}, who used twice-yearly \xmm\ observations
of HD 81809 (G2$+$G9) covering 2001--2007
to reveal a well-defined cycle (presumed to be the G2) 
with quiescent $L_{\rm X}$ varying by a factor of 5 or 6
and matching the 8.2-year period seen in \caii\ HK lines.
%Lastly, \citet{cit:sanz-forcada2013} reported the X-ray detection of 
%a somewhat irregular but convincing 1.6~yr activity cycle 
%with an amplitude of about 50\% in the young solar-type star $\iota$~Hor that had previously been seen in Ca~{\sc ii} H\&K emission \citep{cit:Metcalfe2010}.
Lastly, \citet{cit:sanz2013} reported the X-ray detection of 
a somewhat irregular 1.6~yr activity cycle in the young
solar-type star $\iota$~Hor that had previously been discovered using 
\caii\ HK emission \citep{cit:metcalfe2010}.
X-ray intensity varied by about a factor of two
over the 14 \xmm\ observations that spanned 21 months in
2011--2013.
%Although not all stars are expected to exhibit X-ray cycles
%Although some stars do not exhibit X-ray cycles
Although not all stars will exhibit X-ray cycles
\citep[e.g.,][]{cit:hoffman2012,cit:arlac2014},
%(Hoffman, G\"{u}nther \& Wright 2012),
the above examples illustrate
the potential of detecting cycles using X-ray monitoring.
%even when the monitoring cadence is low.
%
Prox Cen is however, a more challenging case because
it flares more often and its X-ray cycle amplitude appears to 
be smaller than for these other stars.
%The examples above show that a sustained program of X-ray/UV monitoring
%with as few as one or two observations per year is
%sufficient to measure several-year stellar cycles in stars that
%have them, provided that the observations are long enough to
%determine the quiescent emission level and that emission
%varies by factors of a few over the cycle.

In Section~\ref{sec:asas} 
we analyse fifteen years of ASAS optical data,
followed by analysis of four years of \swift\ data 
in Section~\ref{sec:observations}, and then interpretation
of the optical, UV, and X-ray results in Section~\ref{sec:interp}.
%%Section~\ref{sec:otherxray} examines
%% For some reason the above ref/label no longer works and I
%% can't figure out why.  I therefore specify it manually.
Section~5 examines
\swift\ observations in concert with data from other X-ray missions
in order to extend the period of high-energy monitoring,
followed by a summary of results in 
Section \ref{sec:conclusions}.

%%%%%%%%%%%%%%%%%%%%%%%%%%%%%%%%%
\section{Optical Data}
\label{sec:asas}
%%%%%%%%%%%%%%%%%%%%%%%%%%%%%%%%%

% P1
As noted in Section~\ref{sec:intro}, \citet{cit:kiraga2007},
\citet{cit:savanov2012}, and \citet{cit:suarez2016} measured 
rotation periods of around 83 d using ASAS-3 data, in
good agreement with the 83.5 d period measured by 
\citet{cit:benedict1998} using {\it Hubble} data.
\citet{cit:guinan2010}
cites a period of 83.7 d and an activity cycle of $\sim$7.6 yr
derived from an unpublished analysis of ASAS data,
and also says that \rosat, \xmm, and \chandra\ data show a
`corresponding coronal X-ray cycle with an 
expected minimum during 2010/11.'\footnote{
	As explained in Section~\ref{sec:swiftperiods}, 
	we find an X-ray maximum around that time.
}
Using the same ASAS data, \citet{cit:suarez2016} measured
a cycle of $6.8\pm 0.3$ yr.

% P2
For our analysis we downloaded the complete set of ASAS-3
\citep{cit:pojmanski2002}
$V$-band data on Prox Cen 
covering 2000-Dec-27 to 2009-Sep-11
from \url{http://www.astrouw.edu.pl/asas/},
and also added manually processed
data from the ASAS-4 program covering 2010-Jul-08 to 2015-Aug-16.
We used a 4 pixel (1\arcmin) aperture
%% Yes, 1 arcmin.  Pixels are 14.8 arcsec.
(the middle of five available, producing the MAG\_2 measurements)
for photometry
%We used photometry obtained with a 4 pixel (1\arcsec) aperture
%(the middle of five available, producing the MAG\_2 measurements)
as this provided the lowest overall uncertainties.
Of the 1462 measurements with A or B quality flags we kept
only those with magnitudes that fell within 3 standard deviations
of the mean (grouped by observing season, which approximately
coincides with calendar year).
The remaining 1433 observations, typically 3 minutes long,
were made on 1085 nights.
Calibration of the ASAS-4 system, particularly vignetting and PSF,
is not complete, so we used 33 stars 
%near Prox Cen 
to normalize
the ASAS-4 measurements to ASAS-3, with an estimated uncertainty
of around 0.02 or 0.03 magnitudes.  
%The tabulated uncertainty of individual
%observations ranges from 0.025 to 0.068 magnitudes,
%although judging from scatter ASAS-4 data with that nice
%rotational modulation the actual uncerts are considerably less.
Note that since ASAS magnitudes are based on the Tycho-2 system ($V_T$,$B_T$) 
and no color terms were included in ASAS transformation of 
instrumental data, they can differ
slightly from the standard Johnson system, particularly for red stars;
Prox Cen's $V$ magnitude is typically given as around 11.13 
\citep[e.g.,][]{cit:jao2014}.

%% $I$-band data:
%Usefull magnitude ranges:
%in $V$ it is between saturation at 7 (for shorter exposures - 8.) and 14.25
%in $I$ it is 6.5 to 13.25

% P3
We also studied ASAS-3 I-band data, which were less extensive than
for the $V$ band, covering only the 2003, 2005, and 2006 seasons.  
There were 249 measurements on 196 nights with A or B quality,
%$I$ band: 53 double obs's of 249 AB = 196 nights
of which roughly half were collected on the same nights as
% 55 of 152 lines represent 2 obs's per night in or the other band
%  --> 97 common nights
$V$-band observations.  We used the MAG\_3 measurements (5-pixel aperture),
which had the smallest scatter.
%There are also (uncalibrated) I-band data from ASAS-4,
%but these were not included.
%would provide little additional benefit for reasons
%discussed below and were not included.
%but these were not deemed worth the considerable calibration effort
%that would be needed in view of the limited additional information
%that they could provide, as discussed below.

%The measured magnitudes are averages over integrated observations
%(typically 3 minutes)
%and therefore must include some flares.
%While using only quiescent measurements would aid our
%period search, we are unable to reliably identify observations
%affected by flares, and
%Flares have relatively little effect on the total $V$-band intensity,
%however, and
%the number of data points is large enough that
%medium- and long-term trends are still discernible.
% I can't find any info on the uncertainty in ASAS msmts but
% in Kiraga and Stepien's list of periodic stars they list
% a few with amplitudes as small as 8 or 9 mmag.

%While using only quiescent measurements would aid our
%period search, we are unable to reliably identify observations
%affected by flares, and
%the number of data points is large enough and the effect
%of optical flares small enough, that
%medium- and long-term trends are still discernible.

\begin{figure}[t]
%\epsscale{0.50}
\epsscale{1.00}
\plotone{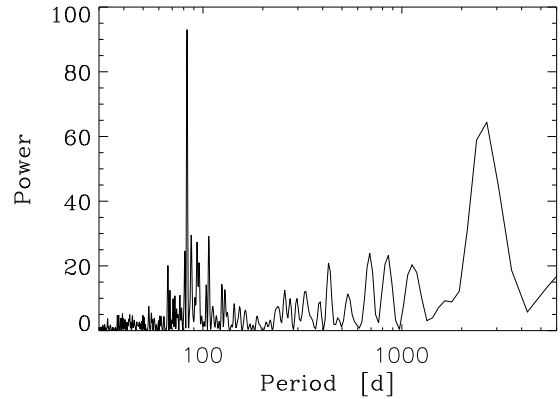}
%Replace 5 lines above with 2 lines below for 1col
%\begin{figure}[t]
%\plottwo{./f1a.eps}{./f1b.eps}
\caption{
Lomb-Scargle periodogram of $V$-band ASAS data.
Rotation peak is at 83.1 d and the broad
peak around 2600 d ($\sim$7 yr) is from the stellar cycle.
}
\label{fig:periodogram}
%\end{figure*}
\end{figure}

% P4
Our search for a rotation period and stellar cycle 
uses a Lomb-Scargle floating mean periodogram analysis 
%(Scargle 1982)
\citep{cit:scargle1982}
with the implicit assumption
that emission nonuniformities 
such as starspots persist for multiple rotation periods and modulate
the observed quiescent emission.  
In the $V$-band data we find two extremely strong peaks 
of $P$ = $83.1\pm0.05$ d 
%and $P$ = $2576\pm52$ d 
and $2576\pm52$ d 
%($7.05\pm0.15$ yr; errors following \citet{cit:baliunas1995}) 
\citep[$7.05\pm0.15$ yr; errors following][]{cit:baliunas1995}
that we interpret as the mean rotational and
magnetic cycle periods, respectively
(see Figure~\ref{fig:periodogram}).
%To give more perspective on the period uncertainties, 
Results when analyzing the ASAS-3 data alone were 82.9 d and 7.91 yr.
%The ASAS-4 data do not span enough
%time to provide a robust cycle period by themselves.
Collectively changing the ASAS-4 measurements by up to $\pm0.05$
magnitudes to gauge the effect of cross calibration uncertainties
barely changed the periods.
For both period determinations, the standard
L-S False Alarm Probability (FAP) $\ll 10^{-20}$
%(**Horne \& Baliunas 1986),  
\citep{cit:horne1986}
although there are many reasons for believing these L-S FAPs overestimate the
certainty of the detections
\citep[e.g.,][]{cit:baliunas1995}.
Monte Carlo simulations for randomly reordered data with
the same time spacing yield FAPs $\sim1\times 10^{-6}$,
so the periods are robust.

\begin{figure}[t]
%\epsscale{0.50}
\epsscale{1.10}
\plotone{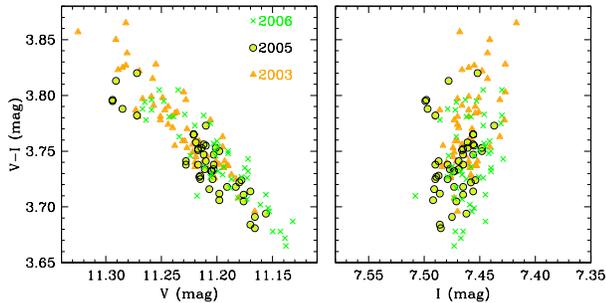}
%Replace 5 lines above with 2 lines below for 1col
%\begin{figure}[t]
%\plottwo{./f1a.eps}{./f1b.eps}
\caption{
Colour diagrams illustrating changes in $V$ over time,
while $I$ remains nearly constant.
%Symbol colour/year convention is (nearly) the same as in
%Figure~\ref{fig:period-rot}, while differentiating
%between data from 2005 and 2006.
}
\label{fig:VIcolors}
%\end{figure*}
\end{figure}

% P5
The smaller set of I-band data yielded a lower-confidence rotation period
of 82.7 d
% with $\sim$2\% peak-to-peak amplitude variations, 
% 2%?? Seems too large.
but spanned too little time to say anything about a multi-year cycle.
%$V$-$I$ colours show a clear decreasing trend with $V$ 
% (Figure~\ref{fig:VIcolors}), 
%thus indicating the star gets brighter in $V$, 
%it also gets less red, suggesting
%that cool starspots are driving the variation.  The lack of a clear $V$-$I$ 
$V$-$I$ colours show a clear trend with $V$ (Figure~\ref{fig:VIcolors})
such that as the star gets brighter it also becomes less red, suggesting
that cool starspots are driving the variation.  The lack of a clear $V$-$I$ 
trend with $I$ suggests that both spots and the quiet photosphere are 
contributing significantly to emission in this band, again underlining that
variation is due to cool features, which are more visible in the red.
%Subsequent discussions of ASAS data refer to $V$-band
%data unless otherwise noted.

%% Jan 26 2016 email from Steve
% > the rotational semi-amp was 0.03
% > the cycle semi-amp was 0.025
% > so the peak-to-peak percentage amplitudes are:
% > [his calcs are wrong.  I calculate them as...
% Amplitudes are given in magnitudes, which are logarithmic
% in base 2.512.  So cycle semi-amp=0.025 mag means p2p
% amp is 0.050 mags --> 2.512^0.050 = 1.047, so about a
% 4.7% p2p difference.
% Mar 22 2016--Now Steve is saying the rot|cycle amps are 0.020|0.0195 mags.
% Those work out to peak2peak Max/min of 1.038 and 1.037.

%%%%%%%%%%%%%%%%%%%%%%%%%%%%
% FIGURE ##
% \label{fig:periodogram}
%%%%%%%%%%%%%%%%%%%%%%%%%%%%
% Use figure* for 2col
%%%%%%%%%%%%%%%%%%%%%%%%%%%%

\begin{figure}[t]
%\epsscale{0.40}
\epsscale{1.15}
%\plotone{./asas_x_rotphase2.eps}
\plotone{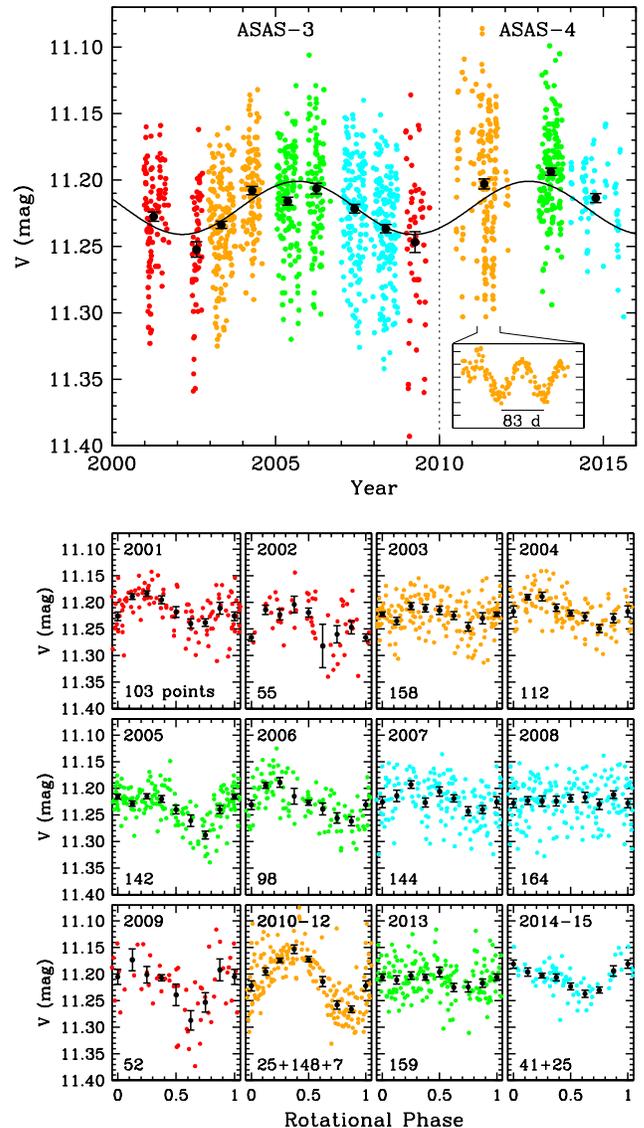}
%Replace 5 lines above with 2 lines below for 1col
%\begin{figure}[t]
%\plottwo{./f1a.eps}{./f1b.eps}
\caption{
Top: ASAS $V$-band data with grouped averages (black)
and best-fit 7.05-yr cycle (sinusoid).
Inset uses same vertical scale.
Bottom: Data are separated by observing season and
phased to a rotation period of 83.1 d,
with 7.05-yr cycle modulation subtracted.
Some years have few measurements and are grouped
with other years.
Colours correspond to associated time intervals in the top panel;
%Data points are coloured to correspond to different years and phases
%of the cycle in the top panel;
black points with error bars are averages over 1/8-period bins.
%Phase shifts ($\Delta\phi$) are relative to the average
%rotational phase.   (Decided not to include phase shifts in plot.)
}
\label{fig:period-rot}
%\end{figure*}
\end{figure}

% P6
To determine the amplitudes of the $V$-band modulations
we first fit and subtracted the 7.05-yr cyclic modulation
(peak-to-peak 0.040 mag = 3.8\%)
and then fit another sinusoid to the residuals
to find the rotational amplitude
(peak-to-peak 0.042 mag = 3.9\%).
Figure~\ref{fig:period-rot} (top panel) plots the data
along with the 7-yr cycle found by the L-S analysis.
To better show the cyclic behavior we also plot
yearly averages with error bars.
% Dates are averages.
(Some of the later years had relatively few measurements
and were grouped together.)
In the bottom panels we separate the data by year,
subtract the 7-yr modulation, and phase all the data
using a common 83.1-d period.
%and fit sines
%to determine the amplitudes and relative phasing.
%{\bf Should Fig 2 include the sines?}
%In the bottom panels we plot 
%data phased to the full-data-set rotational period
%after subtracting cyclic modulation,
%along with averages over 1/8-period bins.

%%%%%%%%%%%%%%%%%%%%%%%%%%%%
% FIGURE ##
% \label{fig:VIcolors}
%%%%%%%%%%%%%%%%%%%%%%%%%%%%
% Use figure* for 2col
%%%%%%%%%%%%%%%%%%%%%%%%%%%%

%%%%%%%%%%%%%%%%%%%%%%%%%%%%
% FIGURE ##
% \label{fig:period-rot}
%%%%%%%%%%%%%%%%%%%%%%%%%%%%
% Use figure* for 2col
%%%%%%%%%%%%%%%%%%%%%%%%%%%%

% P7
Data are colour coded to roughly indicate various phases
of the 7-yr cycle (red for minimum, orange for rising, etc.) but there is
no obvious correlation of rotational phasing or amplitude
with cycle phase.
%Rotational modulation is most pronounced around 2011,
%and the 83-d period can be clearly seen even without
%phasing the data (see inset in top panel).  
%Analysis of each seasonal group
%generally yields rotation periods within a few days of the global
%value of 83.1 days.  
The 2010-12 group has a very well defined 
modulation (see inset in top panel) with a period of 86.3 d.
With the period fixed at 83.1 d, rotational phasing remains remarkably
constant ($\Delta\phi$ between $-0.08$ and $+0.15$)
over nearly the entire 15 years of coverage, despite
significant changes in the modulation amplitude, which falls to
essentially zero in 2008 before recovering.
The apparently stable phasing is consistent with the
findings of \citet{cit:berdyugina2007} that
persistently active longitudes are common in active stars.

%The same rotational
%phasing is roughly maintained for at least 7 years (to 2007),
%with little if any modulation in 2008.  Modulation
%strengthens again (with nearly the same phase as before) 
%starting sometime in 2009 and reaches a peak sometime
%during 2010-12 before fading again by 2013 and 
%modestly strengthening with roughly the same phasing as before in 2014-15.

% P8
If, instead of adopting a fixed average $P_{\rm rot}$, we perform
a period analysis of each time group, we find some evidence for differential
rotation (DR). Including the 2010-12 group noted above with its 86.3-d period,
six intervals have FAP $\leq 10^{-3}$, ranging
from $P_{\rm rot} = 77.1$ days (2001 season; FAP = 8.4$\times
10^{-6}$), to $P_{\rm rot} = 90.1$ days (2009; FAP = 7.0$\times
10^{-4}$; these latter data exhibit modulation with a period of 45 d that
we interpret as arising from nonuniformity on roughly opposite sides 
of the star).
This yields a fractional DR estimate of $\Delta P_{\rm
rot}/\langle P_{\rm rot} \rangle \geq  0.16$, similar to the Sun.
Note that this is a lower limit to DR, since we are likely not
sampling all latitudes on Prox Cen.  Another common DR measure uses
the spread of observed periods: $N \sigma_P/\langle P_{\rm rot}
\rangle$.  For $N$ =3 
\citep[e.g.,][]{cit:lehtinen2016},
%(e.g., Lehtinen et al 2016), 
this measure gives an identical result.

% P9
We believe this is the first DR measurement on such a slowly
rotating, fully convective star.  The estimated $\Delta P_{\rm
rot}$ implies $\Delta \Omega/\Delta \Omega_\odot = 0.33$, which
puts Prox Cen at the edge of the observed $\Delta \Omega - Ro^{-1}$
distribution for single dwarfs 
%(a factor of $\sim$3 below than the overall trend; see Saar 2011, 
%their Fig. 2 left).  
\citep[a factor of $\sim$3 below than the overall trend; see]
[their Fig.~2 left]{cit:saar2011}.
%% Omega = 2*pi/P
The measured DR 
is in better agreement with the $\Delta \Omega - \Omega$ relation
\citep[$\sim$40\% above the trend's extrapolation to lower $\Omega$;]
[their Fig.~2 right]{cit:saar2011}.
We do not observe any correlation between $P_{\rm rot}$
and cycle phase.

% P10
Interestingly, although \citet{cit:hosey2015} included Prox Cen
in their RECONS study of M stars, obtaining 35 nights of
data from 2000--2013 and measuring a standard deviation of 0.0285 mag (2.7\%)
(the 12th largest variability among the 114 stars with
$V$ photometry in their data set),
they did not note any cyclic behavior.
%
%Interestingly, \citet{cit:hosey2015} also studied Prox Cen,
%using 35 nights of RECONS data from 2000--2013,
%and measured a standard deviation of 0.0285 mag (2.7\%),
%the 12th largest variability among the 114 stars with 
%$V$ photometry in their data set,
%but did not note any cyclic behavior.
% May 26 2016, ran stddev.f again on 1433 points
%   11.2187858  0.0425896635
For comparison, the ASAS $V$-band data exhibit a standard deviation 
of 0.0426 mag (4.0\%)
% about the mean $V$ (11.198)
while the Sun's optical intensity varies 
by $\sim$0.1\% over its cycle.
One possibility is that the much sparser RECONS monitoring occurred 
mostly when Prox Cen was near its mean brightness, which
would also explain why their standard deviation is
smaller than ours.
%We also note that \citet{cit:hosey2015} did not see periodicity
%in GJ 581, in contrast to findings by two other groups
%(see Section~\ref{sec:intro}).  (**Does the preceding
%sound snarky?  Is it relevant?)
In any case, the fifteen years of ASAS observations
%, consisting of 1085 nights of observations over fifteen years, 
display highly significant sinusoidal variations
consistent with a 7-yr stellar cycle.
Further interpretation of those results is aided by
analysis of \swift\ X-ray and UV data, which we now discuss.

%Maybe include Figure~\ref{fig:ASASdistribs}
%showing ASAS intensity distributions by year.
%But only if Steve thinks valuable.
%
%%%%%%%%%%%%%%%%%%%%%%%%%%%%%
%% FIGURE ##
%% \label{fig:ASASdistribs}
%%%%%%%%%%%%%%%%%%%%%%%%%%%%%
%% Use figure* for 2col
%\begin{figure}[t]
%\epsscale{0.50}
%\plotone{./shapes.eps}
%\caption{
%ASAS intensity distributions for each year of observations, shown
%in colours.  Black and grey curves show averages for different
%phases of the 7-year cycle.  Hatched region at bottom shows
%differences between distributions for bright and dim phases;
%the larger differences at low brightnesses are discussed in the text.
%** By Steve, I hope **
%}
%\label{fig:ASASdistribs}
%%\end{figure*}
%\end{figure}
%%%%%%%%%%%%%%%%%%%%%%%%%%%%%

%%%%%%%%%%%%%%%%%%%%%%%%%%%%%%%%%%%%%%%%%%%%%%%%%%%%%%%%%%%%%%%%%%%%%%
\section{\swift\ OBSERVATIONS}
\label{sec:observations}
%%%%%%%%%%%%%%%%%%%%%%%%%%%%%%%%%%%%%%%%%%%%%%%%%%%%%%%%%%%%%%%%%%%%%%

The \swift\ satellite is primarily designed to detect and study
gamma ray bursts (GRBs) using its Burst Alert Telescope (BAT)
but also carries an X-Ray Telescope 
\citep[XRT;][]{cit:burrows2005}
and UV/Optical Telescope 
\citep[UVOT;][]{cit:roming2005}
to more accurately determine source positions 
and provide wider spectral coverage.
% of GRBs and their afterglows.
Roughly 20\% of the total observing time is available
to observe non-GRB Targets of Opportunity (TOOs) and other
sources approved in advance through the Guest Investigator (GI) program.
%\swift\ flies in low Earth orbit with a period of $\sim$90 minutes
%and observations are typically broken up into `snapshots'
%several hundred seconds long.

From Apr 2009 through Feb 2013 (\swift\ Cycles 5 to 8),
using a mix of GI time and 
TOO time generously provided by the \swift\ PI, Neil Gehrels,
we obtained 45 XRT and/or UVOT observations of Prox Cen
divided into 125 separate exposures, or `snapshots.'
\swift\ operates in low Earth orbit ($\sim$95-min period) 
and snapshots rarely exceed 2000 s, 
with typical exposures of several hundred seconds.
The XRT was operated in Photon Counting event mode (time-tagged events), 
and data were collected approximately simultaneously with UVOT data.  

%The first 8 observations (21 snapshots; see Table~\ref{table:obslist}) 
%used the UVOT in imaging mode with the UV grism \citep{cit:kuin2015}.
%Most grism snapshots were bracketed by short imaging-mode 
%exposures using the UVW1 and/or UVW2 filters which have
%bandpasses of $\sim$1000~\AA\ centred near 2500 and 1900 \AA,
%respectively.  

The first 8 observations (21 snapshots; see Table~\ref{table:obslist}) 
used the UVOT UV grism \citep{cit:kuin2015} in imaging mode,
% (integrated exposures) 
covering roughly 1700 to 5000 \AA\ in first order.
Resolving power is $\sim$75 at 2600 \AA, and effective area peaks
near the \mgii\ hk blend (2803.5+2796.3 \AA), which is
an analog of the optical \caii\ HK doublet but brighter in M stars.
Most of the grism snapshots were bracketed by short imaging-mode 
exposures using the UVW1 and/or UVW2 filters which have
bandpasses of $\sim$1000~\AA\ centred near 2500 and 1900 \AA,
respectively.  

%In vacuum, MgII k 2796.35
%In vacuum, MgII h 2803.52

% Totals:
% Obs's: 20+3+2+20=45 (2 missing XRT, 1 missing UV/W1)
% Snaps: 55+9+7+54=125
% Ontime: 87041.0 s XRT, 74329.5 s UVOT/W1, 5885.5 s grism

%%%%%%%%%%%%%%%%%%%%%%%%%%%%%%%%
%% Table 1
%% \label{table:obslist}
%%%%%%%%%%%%%%%%%%%%%%%%%%%%%%%%

\begin{deluxetable}{clcDDc}
\tabletypesize{\scriptsize}
%\tabletypesize{\small}
%% Using a larger font causes errors and you'd then have to use longtable.
\tablecaption{\swift\ Observations\label{table:obslist}}
\tablewidth{0pt}
\tablehead{
\colhead{ObsID}	
	& \colhead{Snapshots}	
		& \colhead{Date}
& \multicolumn5c{Exposure Times (s)} \\
	&	& 	& \twocolhead{XRT}
				& \twocolhead{UVW1}
					& \colhead{Grism}
}
%% I'm using the newfangled AAStex6 Decimal alignment scheme,
%% which requires the next line (\decimals), use of D in the
%% deluxetable alignment spec, use of \twocolhead for the
%% affected columns, and attendant increase in the span of
%% multicolumn.  The grism column wouldn't align properly
%% so I used the old padding-with-phn method for that.
\decimals
\startdata
% obslist xrt exposure is apparently before GTI filtering.
% DEADC = 0.996023
% xstats67.woBG says "LiveT" but it's actually ONTIME
% I'm going to list ONTIME
90215002	&502ab	&2009-04-23	&0.0	&158.0	&\phn 833.8	\\
90215003	&503ab	&2009-05-10	&769.7	&153.1	&\phn 294.9	\\
90215005	&505a\tablenotemark{a}bc 
			&2009-05-13	&3111.5	&308.3	&1084.8	\\
90215006	&506ab	&2009-05-27	&2680.3	&222.0 	&1133.8	\\
90215007	&507abc\tablenotemark{a}d	
			&2009-06-19	&2590.0	&222.5	&\phn 733.8 \\
90215008	&508abc	&2009-07-10	&1827.8	&155.8 	&\phn 534.8	\\
90215009	&509abc	&2009-08-01	&1812.7	&150.9 	&\phn 434.8	\\
90215010	&510ab	&2009-08-22	&2019.9	&167.8 	&\phn 834.8	\\
90215011	&511ab	&2009-09-09	&2146.3	&2174.8 &... \\
90215012	&512abc	&2009-10-03	&1840.2	&1873.5 &... \\
90215013	&513abcde &2009-10-23	&2538.4	&2570.9	&... \\
90215014	&514abc	&2009-12-14	&2146.2	&2163.8	&... \\
90215015	&515abc	&2009-12-30	&2232.0	&2322.4	&... \\
90215016	&516ab	&2010-01-19	&0.0	&2025.1	&... \\
90215017	&517a	&2010-01-20	&1846.0	&1857.9	&... \\
90215018	&518abc	&2010-02-04	&2030.9	&0.0 	&... \\
90215019	&519abcde &2010-02-08	&2366.9	&2421.8	&... \\
90215020	&520abc	&2010-03-04	&1847.8	&1872.0	&... \\
90215021	&521ab	&2010-03-24	&1864.8	&1883.9	&... \\
90215022	&522ab	&2010-04-09	&2978.8	&3005.5	&... \\
%  Cycle 5 totals	20 obs's, 55 snaps, 18 XRT, 19 UV/W1, 8 grism
%					38650.2	25710.1	5885.5
31676001	&601abc	&2010-07-10	&1892.1	&1893.6	&... \\
31676002	&602abcd\tablenotemark{a} 
			&2010-12-07	&2894.2	&2994.9	&... \\
31676003	&603ab	&2011-03-12	&3260.5	&3263.6	&... \\
%  Cycle 6 totals	3 obs, 9 snaps	6154.7	6258.5
31676004	&704abcd &2011-09-04	&1989.8	&1987.1	&... \\
31676005	&705abc	&2011-09-08	&776.2	&740.4	&... \\
%  Cycle 7 totals	2 obs, 7 snaps	2766.0	2727.5
31676006	&806ab	&2012-03-30	&1953.2	&1970.8	&... \\
31676007	&807ab	&2012-04-02	&2434.5	&2443.5 	&... \\
31676008	&808abc	&2012-04-06	&2304.3	&2322.2 	&... \\
31676009	&809ab	&2012-04-10	&2557.5	&2563.0 	&... \\
31676010	&810abcde &2012-04-14	&1394.1	&1396.7 	&... \\
91488001	&890abc	&2012-04-18	&2850.8	&2867.9 	&... \\
31676011	&811abcd &2012-04-22	&1027.9	&1032.8 	&... \\
31676012	&812ab	&2012-04-26	&1980.8	&1992.8 	&... \\
31676013	&813ab	&2012-04-30	&2504.8	&2524.7 	&... \\
31676014	&814abc	&2012-05-04	&2534.9	&2545.8 	&... \\
31676015	&815ab	&2012-05-12	&804.9	&814.0 	&... \\
31676016	&816a	&2012-05-13	&847.5	&860.2 	&... \\
31676017	&817abcde &2012-05-16	&2389.5	&2396.7 	&... \\
31676018	&818ab	&2012-05-24	&2103.6	&2115.3 	&... \\
31676019	&819ab	&2012-05-28	&2214.0	&2220.2 	&... \\
31676020	&820abcde &2012-06-03	&1000.4	&1009.0 	&... \\
31676021	&821a	&2012-06-12	&687.0	&704.4 	&... \\
31676022	&822abc	&2012-06-16	&2128.7	&2134.2 	&... \\
91488002	&892a\tablenotemark{a}bc	
			&2012-09-18	&3053.9	&3020.4 	&... \\
91488003	&893ab	&2013-02-18	&2697.8	&2698.7 	&... \\
%  Cycle 8 totals:  20 obs, 54 snaps	39470.1	39633.4
\enddata
\tablecomments{
	Snapshot labels use the leading digit to denote
	the \swift\ observing Cycle, the next two digits for
	the observation number within that Cycle, 
	and abc...\ to indicate
	the snapshots within each observation.
	}
\tablenotetext{a}
	{Snapshot was split into pre-flare and flaring portions.}
\end{deluxetable}
%%%%%%%%%%%%%%%%%%%%%%%%%%%%%%%%
%%% These next two lines are a kludge to force latex to put the
%%% table in a single column---it seems to want to put extra empty
%%% space at the bottom of the table
\vspace{-0.35in}
\mbox{  }

%/data/letg4/bradw/Swift/obslist
%/data/letg4/bradw/Swift/uvstats.out
%/data/letg4/bradw/Swift/xstats67.woBG
%/data/letg4/bradw/Swift/xstats8.woBG

As discussed in Section \ref{sec:grism},
source crowding in the UVOT field (Figure~\ref{fig:UVOTimg})
can lead to overlapping grism spectra and was
a significant problem in Prox Cen's field, which lies roughly 
toward
%in the direction of 
the Galactic centre 
($l$,$b$=313.925,-1.917).
The accompanying short filter exposures did not
suffer this problem so after the first 8 grism observations
we ran subsequent UVOT exposures 
solely using the UVW1 filter in event mode.
%, i.e., with time-tagged events
%as opposed to time-integrated observations such as those with the grism.
%as opposed to the integrated grism observations in imaging mode.

As seen in Table~\ref{table:obslist}, our observations were concentrated in
two time periods.  About 20 observations with an average spacing
of $\sim$18 days were made in Cycle 5, and 18 observations
in Cycle 8 with average 10-day spacing.  
The Cycle 5 observations were primarily designed to look for signs of the
1.2-yr periodicity suggested by \citet{cit:cincu2007},
and the second group focussed on finding rotational variations
with periods of several weeks.
%Those two groups of observations were
%mostly designed to search for a weeks-long rotation period.
Several other observations
were spaced more widely through Cycles 6, 7, and 8 to support multi-year
monitoring.
Total exposures for each Cycle were respectively 
%39 (26 for UV/W1), 6, 3, and 40 ks.
38.7 (25.7 for UVW1), 6.2, 2.7, and 39.5 ks.

%%%%%%%%%%%%%%%%%%%%%%%%%%%%%%%%%%%%%%%%%%%%%%%%%%%%%%%%%%%%%%%%%%%%%%
\subsection{Grism Spectra}
\label{sec:grism}
%%%%%%%%%%%%%%%%%%%%%%%%%%%%%%%%%%%%%%%%%%%%%%%%%%%%%%%%%%%%%%%%%%%%%%

The design and calibration of both UVOT grisms are
thoroughly described by \citet{cit:kuin2015}.
We used the UV grism in `clocked' mode to restrict the
UVOT field of view and block some of the field stars
and their associated spectra.
As mentioned above,
our hope was to use the \mgii\ hk line-to-continuum ratio
as a stellar activity metric but even in clocked mode,
spectra from other stars
often overlapped part or all of the Prox Cen spectrum.
As a result we used the grism for only eight observations
% Obs 502,3,5,6,7,8,9,10
before switching to the UVW1 filter. 
Only four
% 5abc, maybe 7abcd, maybe 8abc, probably 9abc
of the grism observations provided clean \mgii\ hk lines,
and only about half of those had uncontaminated adjoining continuum.
Given this limited data set we
did not expend the considerable effort required to create 
spectra with fully calibrated intensities and wavelengths.

%%%%%%%%%%%%%%%%%%%%%%%%%%%%
% FIGURE
% \label{fig:grism}
%%%%%%%%%%%%%%%%%%%%%%%%%%%%
% Use figure* for 2col
\begin{figure*}[ht!]
\epsscale{1.10}
\plotone{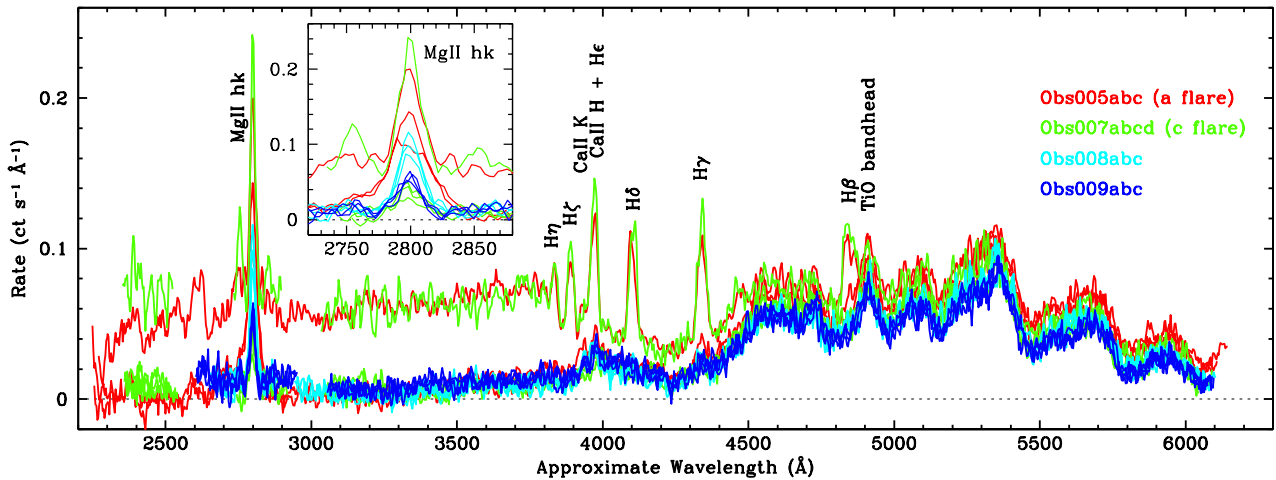}
%Replace 5 lines above with 2 lines below for 1col
%\begin{figure}[t]
%\plottwo{./f1a.eps}{./f1b.eps}
\caption{
Background-subtracted UV grism spectra 
with approximate wavelength calibration,
and detail of \mgii\ emission.
The two higher-rate spectra were obtained during strong flares.
Portions of some spectra are excluded because of
contamination from other sources.
%The broad peak around 4750, 4950, 5150, and 5400 \AA\
%are TiO bandheads.
%Steve's numbers:
%	5166 A
%	5450 A
%	5850 A
%	6158 A
% Left off Obs 2,3,6,10.
Pure 1st order emission extends to $\sim$2700 \AA.
Beyond this, higher orders become increasingly important and
may dominate beyond $\sim$5000 \AA.  Wavelengths also
become more uncertain but the pronounced broad features 
beyond H$\beta$ are mostly due to TiO bandheads.
}
\label{fig:grism}
%\end{figure*}
\end{figure*}
%%%%%%%%%%%%%%%%%%%%%%%%%%%%

We did, however, extract spectra 
%from all but the most contaminated observations 
using the {\tt uvotimgrism}\footnote{
	\url{http://heasarc.gsfc.nasa.gov/docs/software/ftools}
	}
tool, adjusted the spectral and background regions 
to minimize interference from other stars,
and then manually interpolated across contaminated portions of the background
and applied approximate wavelength corrections based
on known spectral features.
%We did, however, extract background-subtracted spectra
%(sometimes using interpolated background) from
%all but the most contaminated observations using
%the {\tt uvotimgrism} tool and then applied ad hoc approximate
%wavelength-dependent dispersion adjustments 
%with appropriately rescaled intensities.
Results for the four observations (thirteen snapshots)
with clean \mgii\ hk lines are shown in Figure~\ref{fig:grism}, 
excluding contaminated regions of each spectrum.
%{\it (I can't (correctly) sum snapshots from a single observation
%because their wavelength bins don't match--isn't worth the
%effort to rebin and then sum.  Or is it?)}
Apart from the prominent \mgii\ hk blend which varies significantly
from one observation to another (see figure inset), 
the quiescent spectra are nearly identical.
%except for the shorter wavelength portions of
%Obs510 which are presumably affected by errors in the background.
%The broad peaks around 4750, 4950, 5150, and 5400 \AA\
%are TiO bandheads.
Note that the \caii\ HK and hydrogen Balmer series lines
commonly seen in K, G, and F stars
are weak or absent in this M star,
except during strong flares (Obs505a and Obs507c) when
the continuum is also enhanced,
thus illustrating why S/N was so low for the HARPS study of Prox Cen 
\citep{cit:gomes2011,cit:gomes2012}, 
which measured the \caii\ HK, \hei\ D3 (5876 \AA),
\nai\ D1 (5890+5896 \AA), and H$\alpha$ (6562 \AA) lines.

%%%%%%%%%%%%%%%%%%%%%%%%%%%%%%%%%%%%%%%%%%%%%%%%%%%%%%%%%%%%%%%%%%%%%%
\subsection{XRT and UVOT Data Extraction}
\label{sec:dataextract}
%%%%%%%%%%%%%%%%%%%%%%%%%%%%%%%%%%%%%%%%%%%%%%%%%%%%%%%%%%%%%%%%%%%%%%

The XRT focuses X-rays onto a $600\times600$-pixel CCD (2.36\arcsec\ pixels)
with a half-power diameter of 18\arcsec.  Prox Cen has a
large proper motion of 3.85\arcsec\ per year so the
expected source position was calculated for each observation in
order to centre the extraction region.
From examining a higher resolution \chandra\ observation (Obs\-ID 49899),
%and after accounting for proper motion and source variability,
we determined that nearby sources unresolved by \swift\ contribute no
more than 1\% of the counts in our 40-pixel-radius (94\arcsec)
source region.  
Point source emission within the 25-times larger background region, an
annulus with radii of 60 and 209 pixels, is similarly unimportant.
All our XRT observations used Photon Counting mode, which has a
time resolution of 2.5 s.
Event pileup is negligible for Prox Cen except during major flares.

%%%%%%%%%%%%%%%%%%%%%%%%%%%%
% FIGURE 2
% \label{fig:UVOTimg}
%%%%%%%%%%%%%%%%%%%%%%%%%%%%
% Use figure* for 2col
\begin{figure}[t]
\epsscale{0.95}
\plotone{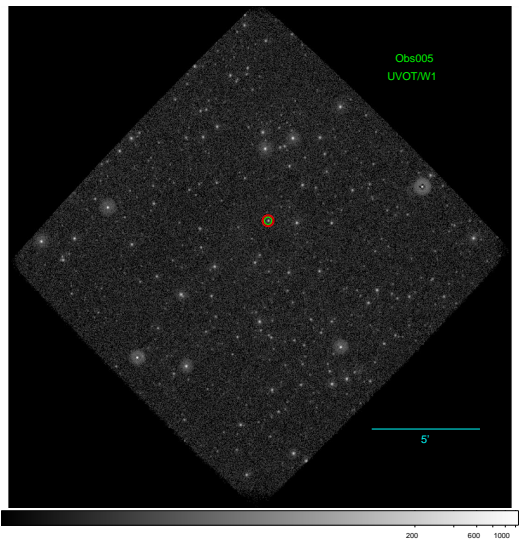}
%Replace 5 lines above with 2 lines below for 1col
%\begin{figure}[t]
%\plottwo{./f1a.eps}{./f1b.eps}
\caption{
Example UVOT image with UVW1 filter.  Source and
background counts were extracted from the green circle
and red annulus, respectively.
}
\label{fig:UVOTimg}
%\end{figure*}
\end{figure}
%%%%%%%%%%%%%%%%%%%%%%%%%%%%

In contrast, the UVOT field is very crowded and some care is
needed in selecting the source and background regions
(see Figure~\ref{fig:UVOTimg}).
Spatial resolution is $\sim$2.5\arcsec\ with 0.502\arcsec\ pixels.
As noted earlier, the grism observations were made in imaging
mode while subsequent observations using the UVW1 filter
were made in event mode, with 0.11 ms time resolution.
We used circular source extraction regions with 10-pixel radii,
and annuli of the same area (radii 13.0--16.4 pixels)
for the background.

There are 45 observations in total, two of them without XRT data and
one missing UVOT data, divided into 125 snapshots
averaging $\sim$720 s each.
After examining light curves, four of the snapshots were split into 
pre-flaring and flaring sections, as noted in Table~\ref{table:obslist}.
%UVOT and XRT not quite simult during event-mode obs's.
%121/(24+20)=2.75 snaps per obs.
Background-subtracted rates for each snapshot were easily calculated
for the UVOT data\footnote{
	All UVOT UVW1 rates account for the
	$\sim$1\% per year decrease in QE reported by
	\citet{cit:breeveld2011}, using mid-2009 as the baseline.
	}
but the XRT analysis was more complicated.  
First, XRT data were divided by energy into a Soft band (0.2--1.2 keV)
and a Hard band (1.2--2.4 keV) with the rationale that the harder
band, although containing fewer counts, is more sensitive to variations
in stellar activity.  The XRT CCD also suffered micrometeoroid damage
early in the mission, leaving some columns and pixels inoperative.
The XRT PSF is broad enough, however, that corrected event rates can
be estimated even when part of the source falls on the damaged regions
by using the `Swift-XRT data products generator'\footnote{
	\url{http://www.swift.ac.uk/user_objects/docs.php}
	}
\citep{cit:evans2009}
which also
applies corrections for event pile-up (only significant during
flares).
% and vignetting (a tiny effect).

% $LEG/Swift/combined.lowest.2014 and .notes:
% 502 no XRT data
% Some short snaps close together treated as one.
% 503/505 (3 days apart)
% 516/517 (516 has no XRT data and the two are <1 day apart
% 518/519 (518 has no UVOT and the two are 4 days apart)
% 704/705 (4 days apart)
% 815/816 (3 days apart)
% 
% Regarding XRT rate corrections for bad columns, vignetting, and pileup:
% "The Swift-XRT data products generator page" is at
% http://www.swift.ac.uk/user_objects/docs.php and says
% "Users may publish any products created using these web tools, provided that 
% this facility is acknowledged. We request this be done both by citing 
% Evans et al. 2009, (MNRAS, 397, 1177) when the data are introduced..., 
% and including the following text in the acknowledgements section of the 
% paper...: This work made use of data supplied by the UK Swift Science Data 
% Centre at the University of Leicester.
% 

%%%%%%%%%%%%%%%%%%%%%%%%%%%%%%%%%%%%%%%%%%%%%%%%%%%%%%%%%%%%%%%%%%%%%%
\subsection{Periodicity Analysis}
\label{sec:swiftperiods}
%%%%%%%%%%%%%%%%%%%%%%%%%%%%%%%%%%%%%%%%%%%%%%%%%%%%%%%%%%%%%%%%%%%%%%

% $LEG/Swift/combined.lowest.2014 and .notes:
% 002 no XRT data
% Some short snaps close together treated as one.
% 003/005 (3 days apart)
% 516/517 (516 has no XRT data and the two are <1 day apart
% 518/519 (518 has no UVOT and the two are 4 days apart)
% 704/705 (4 days apart)
% 815/816 (3 days apart)

%%%%%%%%%%%%%%%%%%%%%%%%%%%%%%%%%
\subsubsection{Selection of quiescent rates}
\label{sec:swiftquiet}
%%%%%%%%%%%%%%%%%%%%%%%%%%%%%%%%%

As noted above, flaring tends to obscure underlying longer term
trends in emission, particularly in the X-ray band where flares can
reach intensities tens or even hundreds of times the quiescent level.
Unlike the G and K stars described in Section~\ref{sec:intro-xrays},
Prox Cen is a relatively active star and determining
when emission is quiescent or flaring is challenging
with limited temporal coverage.  
%Although each \swift\ snapshot is typically only several hundred seconds,
%our observations comprise one to five snapshots spaced at intervals of
%one or more orbital periods of 95 minutes (5700 s).  Most flares evolve over 
%time scales of one hundred to a few thousand seconds, so .
\swift\ snapshots are typically only several hundred seconds long,
too short to tell whether the observed emission is quiescent or flaring.
Each of our observations, however, comprises one to five snapshots spaced at 
intervals of one or more \swift\ orbits ($\sim$95 min), relatively
long compared to typical flare time scales of a few hundred or thousand
seconds, making it much more likely to sample and reliably identify
quiescent periods during a given observation.  
This effort is also aided by having
data in three somewhat independent wavebands: UVW1, and Soft and
Hard X-ray.

\begin{figure}[t]
\epsscale{1.05}
\plotone{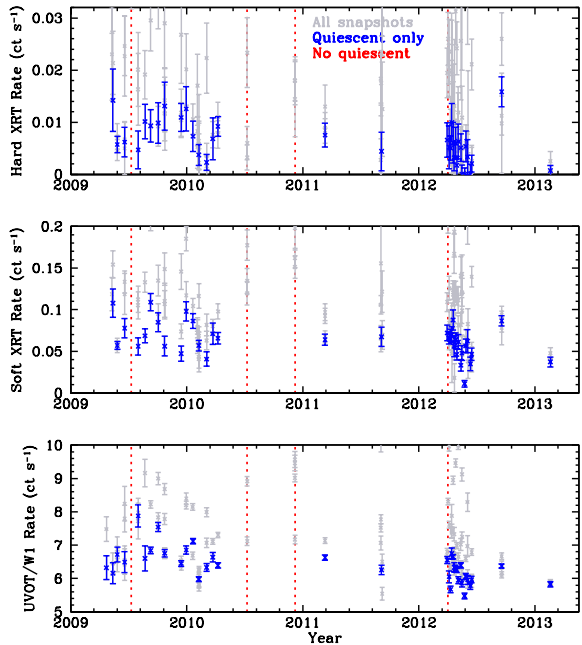}
%Replace 5 lines above with 2 lines below for 1col
%\begin{figure}[t]
%\plottwo{./f1a.eps}{./f1b.eps}
\caption{
\swift\ XRT and UVW1 data, comparing all snapshots' event rates
(grey, excluding roughly a dozen snapshots with bright flares
that are off scale) with quiescent rates (blue).  In four observations,
marked with red vertical lines, all the snapshots were affected
by flares and quiescent rates could not be determined.
}
\label{fig:swiftrates}
%\end{figure*}
\end{figure}

Although multiple wavebands and convenient snapshot spacing help,
there is still the fundamental problem of limited exposure time
and event rates, and
determining `the' quiescent emission level in each band during
an observation remains
a challenge.  After trying several approaches, including
measuring X-ray hardness ratios and various statistical methods,
% non-pedantic summary
we chose a method that, roughly speaking, uses the lowest-rate snapshot
within each observation.\footnote{
        Observations close together in time ($\le$4 d)
        were treated as single observations for this analysis:
        503+505, 516+517, 518+519, 704+705, and 815+816.
        In a few observations, very short snapshots were
        also combined.}
This was simple for the 19 observations in which the lowest-rate
snapshot was the same in all three wavebands.  In 18 other cases
there was no common lowest-rate snapshot and we chose the one 
with the highest significance `lowness,' sometimes averaging rates
from two or even three snapshots if they were very short and/or
their error bars substantially overlapped.
In four observations, emission is decreasing from a prior
flare and does not appear to have reached its quiescent level.  
This was obviously
the case for ObsIDs 508 and 807 and very likely true for 601 and 602,
so they were excluded from further analysis.
%
%we settled on the following criteria:\footnote{
%	Observations close together in time ($\le$4 d)
%	were treated as single observations for this analysis: 
%	503/505, 516/517, 518/519, 704/705, and 815/816.
%	In a few observations, very short snapshots were
%	also combined.}
%\begin{itemize}
%\item[(1)] If the lowest-rate snapshot is the same for all three energy bands, 
%whether or not error bars overlap with those of other snapshots, use that shot.
%\item[(2)] If one or two bands have a common clear lowest-rate snapshot 
%(`clear' means without the error bars overlapping those of other snapshots)
%and the other band(s) have a different lowest-rate shot but with 
%error bars that overlap those of the clear band's shot, for the:
%	\begin{itemize}
%        \item[(a)] clear band(s), use the lowest-rate shot
%        \item[(b)] unclear band(s), average the unclear low with the clear low.
%	\end{itemize}
%\item[(3)] If no band has a clear lowest and (1) does not apply, take the
%average in all bands of the two or three lowest-rate snapshots.
%\item[(4)] If emission is clearly decreasing within all of an
%observation's snapshots then no quiescent rate can be found.  This
%happened for observations 508 and 807, and very likely for 601 and 602.
%%In a few other cases we were able to subdivide snapshots with
%flares into quiescent and flaring intervals.
%\end{itemize}
%%
%Rule \#1 was applied to 19 observations, \#2 to 13, \#3 to 5, and \#4 to 4.
%
%
Figure~\ref{fig:swiftrates}
plots all rate data in grey (with the exception of $\sim$30
off-scale points associated with the largest flares), with our best
estimates of the quiescent
rates shown in blue.  

%%%%%%%%%%%%%%%%%%%%%%%%%%%%
% FIGURE ##
% \label{fig:swiftrates}
%%%%%%%%%%%%%%%%%%%%%%%%%%%%
% Use figure* for 2col
%%%%%%%%%%%%%%%%%%%%%%%%%%%%

%%%%%%%%%%%%%%%%%%%%%%%%%%%%%%%%%
\subsubsection{Evidence for X-ray periodicity}
%%%%%%%%%%%%%%%%%%%%%%%%%%%%%%%%%

Using the \swift\ quiescent-rate data described above we searched for
periodicities using L-S periodograms.  There
are hints of periodicity consistent
with the 7-year photometric cycle, but without a full cycle
% $P_{\rm cyc}$ 
of data the significance is low.  
Comparison of average quiescent rates during Cycles 5 and 8, however,
provides strong evidence for variability on multiyear time scales.
As seen in Table~\ref{table:swiftcycleaves}, there are highly significant
differences in all three energy bands ($12\sigma$ for UVW1,
$6\sigma$ for Soft X-ray, and $5\sigma$ for Hard), 
with the higher rates
occurring as optical brightness nears its minimum.
Relative changes in emission between high and low activity
also follow the expected energy dependent pattern 
(see Section~\ref{sec:intro-xrays}), 
with larger changes observed at higher energies.  

%%%%%%%%%%%%%%%%%%%%%%%%%
\begin{deluxetable}{lccc}
\tabletypesize{\scriptsize}
\tablecaption{Average quiescent event rates 
	(ct s$^{-1}$)\label{table:swiftcycleaves}}
\tablewidth{0pt}
\tablehead{\colhead{Epoch}
                & \colhead{UVW1}
%                        & \colhead{Soft X-ray (0.2-1.2 keV)}
                        & \colhead{Soft (0.2-1.2 keV)}
%                                & \colhead{Hard X-ray (1.2-2.4 keV)}
                                & \colhead{Hard (1.2-2.4 keV)}
}
\startdata
Cycle 5 & $6.597\pm0.030$   &   $0.0662\pm0.0023$   &   $0.00659\pm0.00066$ \\
% Uncorrected numbers below
%Cycle 8 & $5.961\pm0.022$   &   $0.0483\pm0.0018$   &   $0.00244\pm0.00049$ \\
%Difference & $0.628\pm0.037$ &  $0.0179\pm0.0029$   &   $0.00415\pm0.00082$ \\
%Ratio	& $1.1054\pm0.0065$ &	$1.37\pm0.070$	    &   $2.70\pm0.61$ \\
% And with the 1 percent per year (2.87%) QE correction:
Cycle 8 & $6.137\pm0.022$   &   $0.0483\pm0.0018$   &   $0.00244\pm0.00049$ \\
Diff.   & $0.460\pm0.037$   &  $0.0179\pm0.0029$    &   $0.00415\pm0.00082$ \\
Ratio	& $1.075\pm0.006$   &	$1.37\pm0.07$	    &   $2.70\pm0.61$ \\
\enddata
\tablecomments{
	Listed uncertainties are statistical and do not 
	include sys\-tem\-at\-ic uncertainties arising
	from data sampling effects.
        }
\end{deluxetable}
%%%%%%%%%%%%%%%%%%%%%%%%%
%%% These next two lines are a kludge to force latex to put the
%%% table in a single column---it seems to want to put extra empty
%%% space at the bottom of the table
\vspace{-0.25in}
\mbox{  }

\begin{figure}[t]
\epsscale{1.05}
\plotone{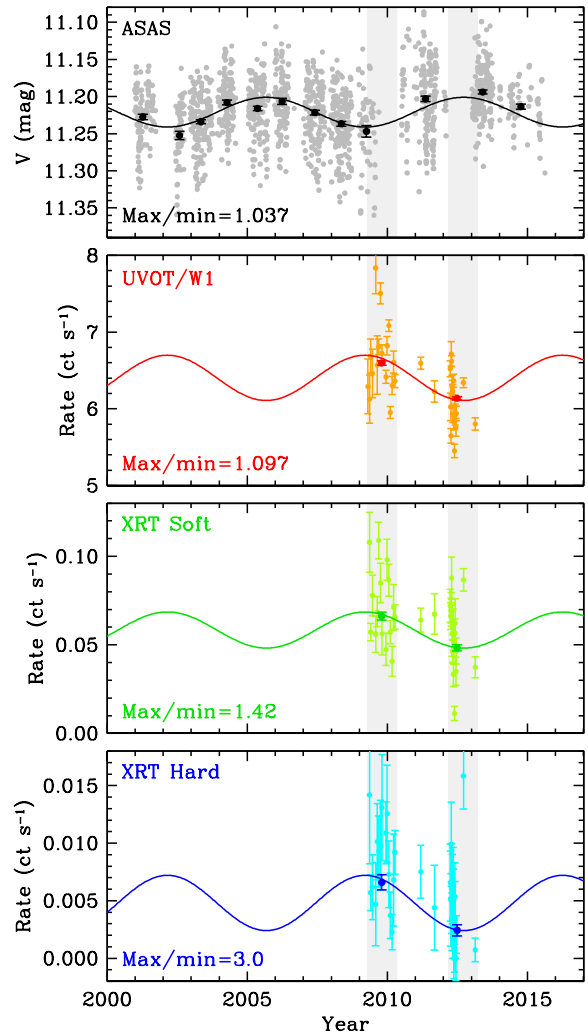}
%Replace 5 lines above with 2 lines below for 1col
%\begin{figure}[t]
%\plottwo{./f1a.eps}{./f1b.eps}
\caption{
ASAS optical photometry and \swift\ quiescent rates.
Lighter shades are used for unbinned points, darker for data
averaged over year-long bins (ASAS) or \swift\ observing Cycles 5 and 8
(vertical grey bands).  
The UV and X-ray sinusoids were fitted using the
period and (inverse) phasing from the ASAS fit.
}
\label{fig:period-cycle}
%\end{figure*}
\end{figure}

Figure~\ref{fig:period-cycle} plots the individual and Cycle-averaged
\swift\ data points along with fitted sinusoids using the same period as
the optical cycle but opposite in phase.
%and scaled to intercept the Cycle-5 and -8 averages.
%This fig now uses Steve's fits, which nearly intercept the ave pts.
Although their uncertainties are relatively large, the
few points from Cycles 6 and 7 generally follow the same curves.
With \swift\ data spanning only about half the optical cycle we
cannot confidently say that there is an X-ray/UV cycle, but the
results are certainly consistent with and highly suggestive of such 
a cycle, as discussed in Section~\ref{sec:interp}.

%%%%%%%%%%%%%%%%%%%%%%%%%%%%
% FIGURE ##
% \label{fig:period-cycle}
%%%%%%%%%%%%%%%%%%%%%%%%%%%%
% Use figure* for 2col
%%%%%%%%%%%%%%%%%%%%%%%%%%%%

%%%%%%%%%%%%%%%%%%%%%%%%%%%%
% FIGURE ##
% \label{fig:XUVrotphase}
%%%%%%%%%%%%%%%%%%%%%%%%%%%%
% Use figure* for 2col
\begin{figure}[t!]
%\epsscale{0.40}
\epsscale{1.05}
\plotone{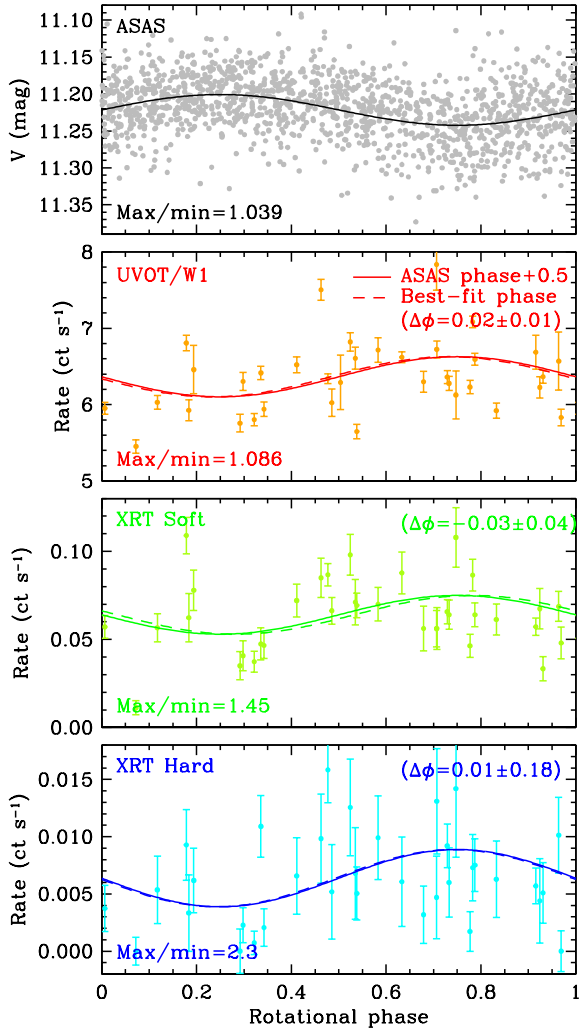}
%Replace 5 lines above with 2 lines below for 1col
%\begin{figure}[t]
%\plottwo{./f1a.eps}{./f1b.eps}
\caption{
ASAS and \swift\ data phased to 83.1-d rotational period.
%Only the 2010--12 ASAS data are used, as they have the best overlap
%with the \swift\ time coverage.
The 7.05-yr fitted cycles from
\protect{Figure~\ref{fig:period-cycle}} have been subtracted
from each data set.
% to better reveal rotational modulation.
Sinusoids were fitted to \swift\ data with phases free
(dashed)
and fixed (solid) to that of the ASAS rotational modulation.
Phase shifts ($\Delta\phi$) are relative to (inverse) ASAS phasing,
i.e., $\Delta\phi=0$ means perfect anti-correlation.
%Max/min ratios refer to the fixed-phase (free-phase) fitted sinusoids.
}
\label{fig:XUVrotphase}
%\end{figure*}
\end{figure}
%%%%%%%%%%%%%%%%%%%%%%%%%%%%

As for rotational periodicity in the \swift\ data, 
the L-S analysis again yields only weak evidence, usually at harmonics
of the ASAS rotational period and a full or half year.
%Figure~\ref{fig:XUVrotphase} plots the UVW1 and Soft X-ray quiescent
%data vs.\ rotation phase, 
%using the same 82.9-d period and opposite phase as the ASAS data
%and after subtracting the 7.05-yr sinusoids plotted 
%in Figure~\ref{fig:period-cycle}.
%The optical anti-phasing is supported by 
%$\chi^2$ fits of the cycle-corrected
%soft X-ray (** and UV?) data to 82.9-d sinusoids
%which yield $\phi = 0.48$ and *TBD*, respectively.
%Fits to the non-corrected data yield similar results, 
%though with larger scatter.  
We note, however, that fitting 83.1-d sinusoids 
%(from the ASAS analysis)
to the UV and X-ray data
after subtracting the 7.05-yr cycle sinusoids fitted
in Figure~\ref{fig:period-cycle}
%yields relative phases of $\phi = 0.40$ and 0.43, ***TBR** respectively
%(see Figure~\ref{fig:XUVrotphase}).  
%That is, rotational modulation of the UV and X-ray emission is nearly
%opposite in phase ($\phi = 0.50$) to the optical emission,
shows that the X-ray/UV rotational modulation is $\sim$exactly out
of phase with the $V$-band variations
(see Figure~\ref{fig:XUVrotphase}).
%a point that will be further discussed in Section~\ref{sec:interp}.
We also note that
the magnitudes of rotational modulation
(maximum/minimum for the fitted sinusoids) 
in the \swift\ energy bands
are very similar to those of the corresponding cyclic modulations 
plotted in Figure~\ref{fig:period-cycle}, just
as the ASAS optical rotational and cyclic modulations are about the same.

%%%%%%%%%%%%%%%%%%%%%%%%%%%%%%%%%
\section{INTERPRETATION OF OPTICAL AND X-RAY/UV PERIODICITY}
\label{sec:interp}
%%%%%%%%%%%%%%%%%%%%%%%%%%%%%%%%%

The simplest interpretation of the above results
is that Prox Cen's X-ray, UV, and optical
intensity variations are all driven by magnetic activity,
with optical intensities anti-correlated with the higher energy emission.
%as indicated in Figure~\ref{fig:period-cycle}.
%Our interpretation of these results is that 
Prox Cen is therefore acting like a
typical `active' FGK star and showing a minimum of magnetic activity 
(and minimum X-ray/UV emission) when it is optically brightest 
\citep[least spotty; e.g.,][]{cit:radick1998,cit:lockwood2007},
%\cite{cit:radick1998}
%\cite{cit:lockwood2007}
unlike the relatively inactive Sun (see also Figure~\ref{fig:VIcolors}).
In these active stars, spots dominate the irradiance changes and
associated active regions (plage) dominate the X-ray emission. 
\citep[Note that spot umbrae themselves are not typically very 
bright in X rays;][]{cit:sams1992}.
%Sams, Golub \& Weiss 1992.) 
%\cite{cit:sams1992}

This situation may extend to late M
dwarfs as well; despite being old, Prox Cen has a relatively high
$L_{\rm X}/L_{bol}$, and is therefore still `active.'
Along these lines, the relatively small photometric amplitudes
seen here may actually imply more significant spot area variations.
We used BT-Settl models \citep{cit:allard2012} to model the photometry
and find reasonable results that roughly match Prox Cen's variations 
($\Delta(V-I) \approx 0.18$, $\Delta V \approx 0.15$, $\Delta I \approx 0.03$;
see Figure~\ref{fig:VIcolors}) for a range of parameters. 
Generally the modelled change in spot 
filling factor $\Delta f_S$ is in the range 0.05 - 0.10
on top of a significant level of baseline coverage (total $f_S > 20$\%).

In comparison to the six stars with measured X-ray stellar cycles
(see Section~\ref{sec:intro})
the cycle amplitude of Prox Cen in X rays is relatively small,
with $L^{\rm max}_{\rm X}/L^{\rm min}_{\rm X}$ roughly 1.5 
versus 2--6 for the G and K stars
(see Table~\ref{table:6stars}).
%depending on the 
%energy range (see Figure~\ref{fig:period-cycle}) vs.\ ratios
%(see Table~\ref{table:6stars})
%of roughly 2 for 61 Cyg A 
%\citep[K5V;][]{cit:hempelmann2006},
%3 for $\alpha$ Cen B 
%\citep[K1V;][]{cit:ayres2009},  
%7 for HD 81809 
%\citep[G2+G9;][]{cit:favata2008},
%and 6 for the Sun 
%\citep[G2V;][]{cit:judge2003}.  
Prox Cen is, however, the most active star
in this group with $\log(L_X/L_{\rm bol}) \sim -4.4$,
and there seems to be a general trend towards lower fractional
quiescent variability amplitudes as activity increases.
%Indeed, the most extensive study to date of X-ray and EUV emission of an active star---the RS~CVn-type binary AR~Lac , which has a ``saturated''
%$\log(L_X/L_{\rm bol}) \sim -3$---failed to find evidence for cyclic behavior in data obtained over a period of 33 years, and instead found that deviations from a constant emission level typically amounted to only 15\% \citep{cit:drake2014}.

%%%%%%%%%%%%%%%%%%%%%%%%%%%%
% FIGURE ##
% \label{fig:rossbyetc}
%%%%%%%%%%%%%%%%%%%%%%%%%%%%
% Use figure* for 2col
\begin{figure}[h]
\epsscale{1.10}
\plotone{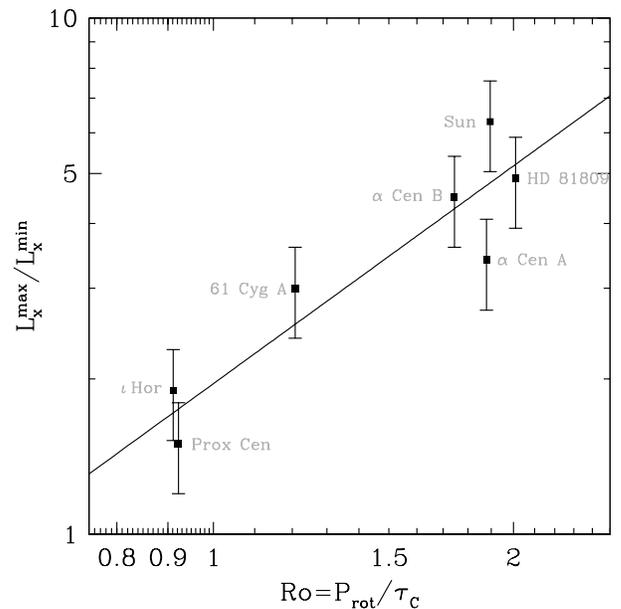}
\caption{X-ray cycle amplitude versus Rossby number,
using data from Table~\ref{table:6stars}.
The fitted power law is 
$L^{\rm max}_{\rm X}/L^{\rm min}_{\rm X}=1.97Ro^{1.39}$.
Cycle amplitudes can vary (particularly for $\iota$ Hor)
so uncertainties are not
well determined; $\pm20$\% error bars are shown for illustrative purposes.
}
\label{fig:rossbyetc}
\end{figure}
%%%%%%%%%%%%%%%%%%%%%%%%%%%%

To investigate this further, we compared various stellar
parameters such as mass and rotation period
and found that the best correlation was between
X-ray luminosity changes and
Rossby number $Ro = P_{\rm rot}/\tau_C$,
as shown in Figure~\ref{fig:rossbyetc}.
The best fit using a power law yields
$L^{\rm max}_{\rm X}/L^{\rm min}_{\rm X} \propto Ro^{1.4}$,  
%Lxmax/Lxmin ~ 1.97 (tauC/Prot)^(1.39)
which is reminiscent of the well known rotation-activity relationship
$L_{\rm X}/L_{\rm bol} \propto  Ro^{-2.7}$ for partially convective
stars below the saturation regime \citet{cit:wright2011},
which \citet{cit:wright2016} showed also applies to Prox Cen and
three other fully convective stars.
Note that the similar characteristics of Prox Cen
and $\iota$ Hor in terms of 
$L^{\rm max}_{\rm X}/L^{\rm min}_{\rm X}$,
$L_{\rm X}/L_{\rm bol}$,
and $Ro$ 
despite their vastly different 
masses and rotation periods
underlines the importance of both rotation {\em and} convective time scales, 
so that the `rotation-activity' relationship is more properly thought of
as the `Rossby-number/activity' relationship.

In any case, the limited available data suggest that 
below the saturated regime ($L_{\rm X}/L_{\rm bol} \la -3$), 
smaller Rossby number means 
higher coronal activity and lower X-ray cycle contrast,
%In any case, the limited available data suggest X-ray cycle
%amplitudes increase with Rossby number while coronal emission decreases,
likely because more active stars are 
more covered with X-ray emitting active regions even at their cycle minima,
so that the contrast over a cycle is lower than for less active stars. 
This may be due to 
modulated or overlapping cycles (i.e., multiplicative or
additive cycles), perhaps in combination with a steady level of
underlying activity generated by, e.g., a non-cycling turbulent dynamo.
Prox Cen shows no obvious signs of multiple cycles
but \citet{cit:sanz2013} suggest that a second longer cycle
that modulates the 1.6-yr cycle might account for some of the
irregular behavior of $\iota$ Hor.
%but \citet{cit:sanz2013} suggest that $\iota$ Hor's
%somewhat irregular behavior might be caused by a second longer cycle
%that modulates its 1.6-yr cycle.

%multiple concurrent cycles (not obvious in Prox Cen
%but possibly in $\iota$ Hor******), 
%overlapping cycles (the extended solar cycle writ larger),
%and/or a steady activity 
%background generated by, e.g., a non-cycling turbulent dynamo.  
%The latter two possibilities seem most likely for Prox Cen. 

%% Force a page break so Table 3 doesn't get broken.
%\newpage

%%%%%%%%%%%%%%%%%%%%%%%%%
\begin{deluxetable*}{lccccccc}
\tabletypesize{\scriptsize}
\tablecaption{Stellar parameters\label{table:6stars}}
\tablewidth{0pt}
\tablehead{\colhead{Star}
		& \colhead{Type}
		& \colhead{$M/M_{\sun}$}
                & \colhead{$L_{\rm X}/L_{bol}$\tablenotemark{a}}
                        & \colhead{$L^{\rm max}_{\rm X}/L^{\rm min}_{\rm X}$}
                                & \colhead{$P_{\rm cyc}$ (yr)}
						& \colhead{$P_{\rm rot}$ (d)}
						& \colhead{$\tau_C$\tablenotemark{b} (d)}
}
\startdata
Prox Cen     	& M5.5V	& 0.12	& -4.4  & 1.5	& 7.1 	& 83  
						& 90	\\
61 Cyg A     	& K5V	& 0.70	& -5.6\tablenotemark{c}  
				& 2.8\tablenotemark{c}  	
					& 7.3\tablenotemark{d}  
							& 35.4\tablenotemark{e}    
								& 29.3	\\
$\alpha$ Cen B 	& K1V	& 0.91	& -6.1\tablenotemark{f}
				& 4.5\tablenotemark{f}	
					& 8.1\tablenotemark{f}   
						& 37\tablenotemark{c}    
								& 21.2	\\
$\alpha$ Cen A 	& G2V 	& 1.1	& -7.1\tablenotemark{f}  
				& $\sim$3.4\tablenotemark{f} 	&$\sim$19\tablenotemark{f} 
						& 28\tablenotemark{f}   & 14.9	\\
Sun          	& G2V	& 1.0	& -6.7  & 6.3\tablenotemark{g}  	
					& 11    & 25.4  & 13.4	\\
HD 81809       	& G2V
			& 1.7
			& -5.9\tablenotemark{h}  
				& 5\tablenotemark{h}  	
					& 8.2\tablenotemark{d}  
						& 40.2\tablenotemark{e}
						& 20.0	\\
$\iota$ Hor 	& F8V
			& 1.25
			& -5.0\tablenotemark{i}  
				& $\sim$1.9\tablenotemark{i}  	
					& 1.6\tablenotemark{i}  
						& 8.2\tablenotemark{j}
						& 9.0	\\
\enddata
\tablenotetext{a}
	{$L_{\rm X}/L_{bol}$ is computed using the average of 
	$L^{\rm max}_{\rm X}$ and $L^{\rm min}_{\rm X}$ over 0.2--2 keV.}
%%% Negative vspaces can be used to tweak where  things end up
%\vspace{-0.05in}
\tablenotetext{b}
	{Convective turnover times ($\tau_{C}$) are taken from
	\citet{cit:gunn1998}
	with extension to M dwarfs following
	\citet{cit:gilliland1986}.
%Values for Prox Cen and 
%the slightly evolved HD 81809 are less certain.
	}
%\vspace{-0.05in}
\tablenotetext{c}
	{From \citet{cit:robrade2012}.}
%\vspace{-0.05in}
\tablenotetext{d}
	{From \citet{cit:baliunas1995}.}
%\vspace{-0.05in}
\tablenotetext{e}
	{From \citet{cit:donahue1996}.}
%\vspace{-0.05in}
\tablenotetext{f}
	{From \citet{cit:ayres2014}.  \citet{cit:robrade2012} estimate
	$L^{\rm max}_{\rm X}/L^{\rm min}_{\rm X}\sim10$ for $\alpha$ Cen A 
	but the A and B components are
	not well resolved by \xmm\ and there are also concerns regarding 
	low-energy calibration.}
%\vspace{-0.05in}
\tablenotetext{g}
	{From \citet{cit:judge2003}.}
%\tablenotetext{?}
%	{HD 81809 is somewhat evolved, 
%	with M$\sim1.7M_{\sun}$ \citep{cit:pourbaix2000} (redundant?)
%	and R$\sim3 R_{\sun}$ ** REF?.}
%\vspace{-0.05in}
\tablenotetext{h}
	{From \citet{cit:favata2008},  
%	We exclude one anomalously bright and high-$T$ measurement
%	that the authors suggest was affected by a flare.}
	excluding the anomalous measurement likely affected by a flare.}
%\tablenotetext{?}
%	{From \citet{cit:pourbaix2000}.}
%\tablenotetext{?}
%	{i Hor type and mass from *****Vauclair et al 2008}
\tablenotetext{i}
	{From \citet{cit:sanz2013}.}
\tablenotetext{j}
	{We use an average of values ranging between 7.9 and 8.6 d
	found by \citet{cit:saarosten1997}, \citet{cit:saaretal1997},
	\citet{cit:metcalfe2010}, and \citet{cit:boisse2011}.}
\end{deluxetable*}
%%%%%%%%%%%%%%%%%%%%%%%%%

%%%%%%%%%%%%%%%%%%%%%%%%%%%%%%%%%%%%%%%%%%%%%%%%%%%%%%%%%%%%%%%%%%%%%%
\section{OTHER X-RAY OBSERVATIONS}

\label{sec:otherxray}
%%%%%%%%%%%%%%%%%%%%%%%%%%%%%%%%%%%%%%%%%%%%%%%%%%%%%%%%%%%%%%%%%%%%%%

%Prox Cen has been observed by almost every X-ray observatory to date
% (exception is Suzaku)
%and there have been suggestions of an emission cycle by 
%Haisch et al.\ (1990),
%but as was the case with $\alpha$ Cen,
%the lack of sustained monitoring with a single set of instruments
%makes drawing any conclusions difficult.
%Also, many of the archival
%X-ray observations are too short to tell if the star was flaring,
%which is much more common on M stars than on K and G stars
%(and more common on mid-type M's than early M's.)
%[And IUE spectra are, I think, integrated exposures so it's hard
%to tell if there was any flaring.]
%On the other hand, flares generally erupt from active regions
%which generate nonuniform quiescent emission across the surface
%of the star, which in turn may modulate the observed emission
%as the star rotates and provide a means to measure the rotational
%period even when the star is rotating too slowly to measure
%Doppler shifts in emission lines.  ** Explain more about
%how rotation periods are measured, here or somewhere. **

Although \swift's monitoring of Prox Cen is in some respects
the most extensive
of any X-ray mission to date, it covers only about half of
the proposed 7-yr cycle and several other missions have comparable
or greater total exposure time, often at higher event rates.
As noted in Section~\ref{sec:intro-xrays},
there are complications in comparing data from multiple instruments,
and as seen in Section~\ref{sec:swiftquiet},
determination of the `true' quiescent emission level during
a given epoch may not
be possible using a single observation,
but thoroughness demands that we try to incorporate data 
from other missions in our study.
%We have therefore studied data from
%other missions back to 1994, effectively the beginning of modern
%X-ray astronomy using CCD technology.

The first pointed X-ray observations (as opposed to survey scans) of Prox Cen
were made by \einstein\ in 1979 and 1980 and \exosat\ in 1985,
using proportional counters.
Excluding very brief observations, the {\it R\"{o}ntgen Satellite} (\rosat) 
Position Sensitive Proportional Counter (PSPC) collected $\sim$36 ks
of data during four observations in 1993 and early 1994.
%The sparse temporal coverage of these observations and the
%limitations of proportional counters
The {\it Rossi X-Ray Timing Explorer (RXTE)} made two sets of
observations in 1996 Feb (51 ks) and 2000 May (45 ks),
% providing a potentially useful internal comparison of rates 
% over an interesting interval of time, 
but its proportional counters 
have very little effective area below 2 keV,
spatial resolution is poor ($1^\circ$ full width at half maximum intensity, 
encompassing other sources), 
and the background is several times as large as the quiescent
signal from Prox Cen and difficult to model.
See \citet{cit:guedel2002} and references therein for details
regarding X-ray observations prior to 2002.

Given the sparse temporal coverage of these earlier missions,
the limited energy resolution of proportional counters, and
significant cross-calibration uncertainties we restrict our analyses 
to missions with CCD detectors
and list those observations in Table~\ref{table:otherXray}.
\chandra\ data were taken from its data archive\footnote{
	\url{http://cxc.harvard.edu/cda/}.
	}
while other data were downloaded from the 
High Energy Astrophysics Science Archive Research Center.\footnote{
	\url{http://heasarc.gsfc.nasa.gov/docs/archive.html}.
	}
%Our approach is based on the idea that sufficiently long and/or
%well spaced observations
%provide a representative sample of stellar behavior, particularly
%at the quiescent emission levels that are of interest here.
%These criteria are admittedly vague but it is obvious that more
%exposure time is better and 
We also reexamine the \swift\ data,
this time treating data from Cycle 5
collectively,
%(totaling 38.7 ks), 
and likewise for Cycle 8.
%(39.5 ks).  
The handful of observations from Cycles 6 and 7
are not included as they do not provide an
adequate data sample for this analysis.

%%%%%%%%%%%%%%%%%%%%%%%%%
\begin{deluxetable}{ccccD}
\tabletypesize{\tiny}
\tablecaption{Cumulative \swift\ and other X-ray observations
	\label{table:otherXray}}
\tablewidth{0pt}
\tablehead{\colhead{Mission}
                & \colhead{Instrument}
                        & \colhead{ObsID}
                        	& \colhead{Date}
                                	& \twocolhead{Exp.~(ks)}
}
\decimals
\startdata
\asca 		& SIS	& 21022		& 1994-03-19	& 28.3	\\
\asca 		& SIS	& 27027		& 1999-08-22	& 57.9	\\
\chandra 	& ACIS	& 49899$+$641	& 2000-05-7,9	& 48.9	\\
\chandra 	& HETG	& 2388		& 2001-09-13	& 42.9	\\
\chandra 	& HETG	& 12360		& 2010-12-13	& 79.3	\\
\xmm 		& PN	& 49350101	& 2001-08-12	& 67.4	\\
\xmm 		& PN	& 551120[3,2,4]01	& 2009-03-10,12,14	& 88.8	\\
\swift 		& XRT,UVW1	& Cycle 5	& 2009-10-25\tablenotemark{a}	& 38.7	\\
\swift 		& XRT,UVW1	& Cycle 8	& 2012-09-08\tablenotemark{a}	& 39.5	\\
\chandra 	& HRC-I	& 14276		& 2012-06-15	& 49.6	\\
\chandra 	& HRC-I	& 17377		& 2015-12-09	& 35.9\tablenotemark{b}	\\
\enddata
\tablecomments{
	Exposure times are durations, without deadtime corrections.
	}
\tablenotetext{a}{Midpoint of Prox Cen observations 
	for that \swift\ observing Cycle.}
\tablenotetext{b}{Excludes 13.8 ks when telemetry was saturated.}
\end{deluxetable}
%%%%%%%%%%%%%%%%%%%%%%%%%
%%% These next two lines are a kludge to force latex to put the
%%% table in a single column---it seems to want to put extra empty
%%% space at the bottom of the table
%\vspace{-0.25in}
%\mbox{  }

\begin{figure}[t]
\epsscale{1.15}
\plotone{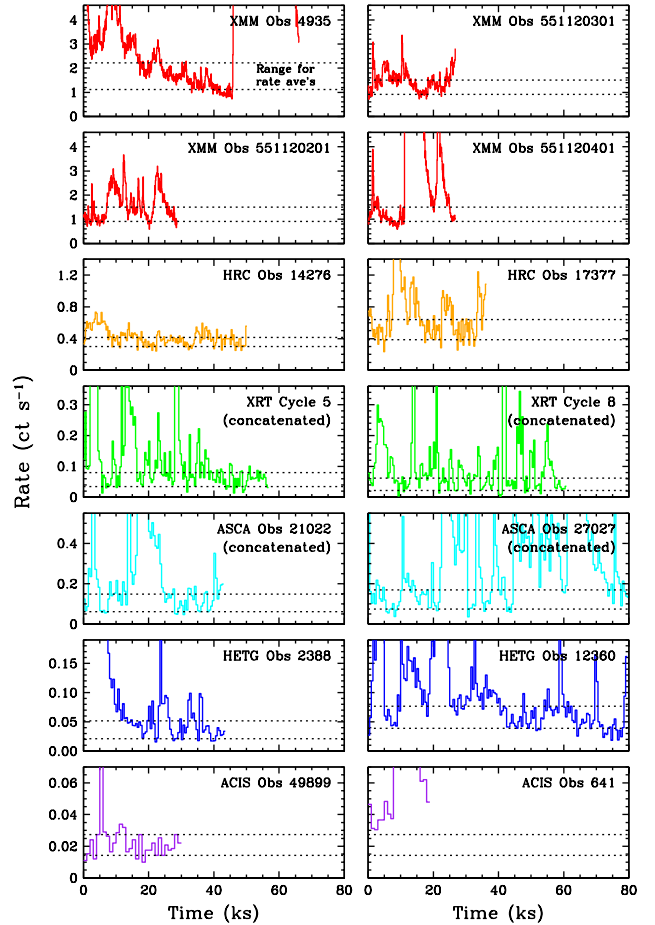}
\caption{
Light curves from X-ray missions, using data between 0.5 and 2.5 keV
(0.6--2.5 keV for \asca, full range for \chandra\ HRC-I).  
%Bin sizes range from 100 to 1000 s (top to bottom) and 
Time gaps in \swift\ XRT and \asca\ SIS
data have been removed for clarity.  Horizontal dotted lines mark the 10 and 60
percentile quiescent rates shown in Figure~\ref{fig:ratedistrib}.
It is likely that the first \xmm\ and second HETG observations 
sampled little if any quiescent emission.  
%The quiescent sampling
%range for the second \asca\ observation may also be too high.
}
\label{fig:lcALL}
%\end{figure*}
\end{figure}

Including the \swift\ XRT and UVOT data, we have a total of
13 data sets from 7 instruments; all but one instrument has
two epochs of data.
We make background-subtracted light curves for each data set 
with bin sizes ranging from 100 to 1000 s, 
depending on source and background rates.  
Source extraction regions are chosen to enclose $\sim$95\% of source
counts and livetime fractions are $\sim$99\%, with noted exceptions.
To reduce the effect of cross-calibration uncertainties
we use a common energy range of 0.5--2.5 keV unless otherwise
noted.  Before explaining how the light curves were used to 
determine quiescent emission levels 
we briefly describe the data from each instrument,
proceeding in roughly chronological order.
Background-subtracted light curves with corrections for 
enclosed energy fraction, vignetting, and livetime 
are shown in Figure~\ref{fig:lcALL}.

%%%%%%%%%%%%%%%%%%%%%%%%%%%%
% FIGURE ##
% \label{fig:lcALL}
%%%%%%%%%%%%%%%%%%%%%%%%%%%%
% Use figure* for 2col
%%%%%%%%%%%%%%%%%%%%%%%%%%%%

%in order to obtain at least 20 counts in each bin.
%XRT cycle 8 sometimes has fewer but we don't use those for 10-60% calcs.

%%%%%%%%%%%%%%%%%%%%%%%%%%%%%%%%%%%%%%%%%%%%%%%%%%%%%%%%%%%%%%%%%%%%%%%
%\subsection{\rosat}
%\label{sec:otherxray-rosat}
%%%%%%%%%%%%%%%%%%%%%%%%%%%%%%%%%%%%%%%%%%%%%%%%%%%%%%%%%%%%%%%%%%%%%%%
%
%I might look at these data to bolster the argument that a
%single observation during an epoch probably isn't adequate
%(because of rotational effects or whatever).
%All 4 \rosat\ obs's are roughly 1 epoch--they span 13 months--and 
%I suspect they have a wide range of `quiescent' rates.

%%%%%%%%%%%%%%%%%%%%%%%%%%%%%%%%%%%%%%%%%%%%%%%%%%%%%%%%%%%%%%%%%%%%%%
\subsection{\asca}
\label{sec:otherxray-asca}
%%%%%%%%%%%%%%%%%%%%%%%%%%%%%%%%%%%%%%%%%%%%%%%%%%%%%%%%%%%%%%%%%%%%%%

The {\it Advanced Satellite for Cosmology and Astrophysics} (\asca) 
made two observations in 1994 Mar and 1999 Aug.  Measurements
spanned roughly 1.5 and 2 days, respectively, but the exposures were
separated into many segments of a few ks each with similar length breaks
between them. 
%and the net exposure time is roughly one-third of that.  
We have analyzed only the Solid-state Imaging Detector
(SIS) data as they have
better low-energy efficiency and energy resolution than the 
Gas Imaging Spectrometers (GIS), and have excluded a small amount of
low-bit-rate data that suffer from telemetry saturation.
SIS data from both detectors (SIS0 and SIS1) were collected
in 1-CCD mode (except for parts of the first observation, which used
2-CCD mode), and were processed uniformly as `Bright' mode data using
standard event screening.  We extracted source data 
from SIS0/chip1 and SIS1/chip3 using circles of radius 3\arcmin,
or slightly elliptical regions of the same area when the source was
too close to the chip edge to fit a circle.  
Background was collected from a narrow
ellipse of the same area along the outer edge of the chip, and the
net enclosed energy fraction of the source region is $\sim$0.69.
Pileup is never a concern given the broad instrumental PSF.

Calibration uncertainties with this early CCD mission are
large, particularly for data taken after 1994 and at low energies.\footnote{
	\url{https://heasarc.gsfc.nasa.gov/docs/asca/cal\_probs.html}
	}
Examples include unphysical spectral features below 0.6 keV and
a significant but uncalibrated decrease in effective area 
below $\sim$1 keV over time.
To reduce the impact of these issues we extract data from 0.6 to 2.5 keV
instead of the usual 0.5--2.5 keV, but the \asca\ results must be
viewed with skepticism.

%%%%%%%%%%%%%%%%%%%%%%%%%%%%%%%%%%%%%%%%%%%%%%%%%%%%%%%%%%%%%%%%%%%%%%
\subsection{\chandra\ ACIS}
\label{sec:otherxray-acis}
%%%%%%%%%%%%%%%%%%%%%%%%%%%%%%%%%%%%%%%%%%%%%%%%%%%%%%%%%%%%%%%%%%%%%%

\chandra\ made two observations in 2000 May using ACIS-S3.
The intention was to use ACIS with the LETG transmission grating but a hardware
failure prevented its insertion.  The core of the source was therefore
heavily piled up and produced a prominent CCD readout transfer
streak.  We analysed these data following procedures described in
\citet{cit:wargelin2002}, extracting unpiled data from the readout
streak and annuli around the source, using regions listed in Table 1
of that work.  We then created light curves (1000-s bins), 
subtracted background,
and rescaled the net rates for each bin to recover the event rates
that would have been obtained if there was no pile-up.
Scaling factors ranged from 0.041 to 0.060, i.e. the
measured rates were only $\sim$4--6\% of the unpiled rates.
To better indicate the measured event rates we
scaled everything back down with a common factor of 0.05 before plotting
in Figures~\ref{fig:lcALL} and \ref{fig:ratedistrib}.

%%%%%%%%%%%%%%%%%%%%%%%%%%%%%%%%%%%%%%%%%%%i%%%%%%%%%%%%%%%%%%%%%%%%%%%
\subsection{\chandra\ HETG}
\label{sec:otherxray-hetg}
%%%%%%%%%%%%%%%%%%%%%%%%%%%%%%%%%%%%%%%%%%%%%%%%%%%%%%%%%%%%%%%%%%%%%%

The two High Energy Transmission Grating (HETG) measurements 
were made in 2001 and 2010, both near expected cycle maxima.
The event rate for 0th order is rather low so we also
included $\pm$1st orders, applying the standard spectral
and background extraction regions and wavelength dependent filtering.
%to the dispersed orders and associated background regions.  
Pileup reached several percent during a few flares
but this does not affect our quiescent emission analysis.

%%%%%%%%%%%%%%%%%%%%%%%%%%%%%%%%%%%%%%%%%%%%%%%%%%%%%%%%%%%%%%%%%%%%%%
\subsection{\xmm}
\label{sec:otherxray-xmm}
%%%%%%%%%%%%%%%%%%%%%%%%%%%%%%%%%%%%%%%%%%%%%%%%%%%%%%%%%%%%%%%%%%%%%%

% for instrument deadtime, which was generally $\sim$1\%
% except for XMM PN ObsIDs 4935 (30\% deadtime) and 55112 (5\%).

Like the \chandra\ HETG measurements, the \xmm\ observations were 
%unfortunately 
both made near cycle maxima.  Event rates were
high enough with the EPIC PN detector that we did not include data
from the lower-rate MOS detectors or RGS gratings. 
ObsID 4935 was made using PN small window mode (5.7 ms frame time)
and pileup was always negligible,
although the deadtime fraction was 30\%.
%	\footnote{
%	See 
%{\url http://xmm-tools.cosmos.esa.int/external/xmm\_user\_support/documentation/uhb/pileupcomp.html},
%Figure 116 of the `XMM-Newton Users' Handbook,'
%for a plot of pileup fraction vs.\ event rate
%comparing \chandra\ ACIS, XMM PN, and XMM MOS.
%	}
ObsID 55112 used large window mode (47.7 ms frame time)
with 5\% deadtime
and pileup was less than 1\% except during large flares.
%which are unimportant for the quiescent emission analysis.
The enclosed energy fraction of the 25\arcsec-radius source
region is 0.70.
% in both cases.
%Note that the lower-rate MOS data {\em are}
%used in the microflaring analysis of Section *** (NO!!!) because of
%their minimal pileup.

%%%%%%%%%%%%%%%%%%%%%%%%%%%%%%%%%%%%%%%%%%%%%%%%%%%%%%%%%%%%%%%%%%%%%%
\subsection{\swift\ XRT and UVOT}
\label{sec:otherxray-swift}
%%%%%%%%%%%%%%%%%%%%%%%%%%%%%%%%%%%%%%%%%%%%%%%%%%%%%%%%%%%%%%%%%%%%%%

The cumulative exposures for \swift\ Cycle 5 data (2009 Apr--2010 Apr)
are 39 ks for the XRT and 26 ks for UVOT/UVW1, and 
$\sim$39.5 ks for both instruments during Cycle 8 (2012 Mar--2013 Feb).  
Pileup was negligible except during large flares.
We include UVOT data in this analysis mostly to illustrate
differences in the rate distributions of UV and X-ray emission.
%The total Cycle 6 and 7 exposure was 11 ks (and usually flaring),
%probably too little to add to what can be seen 
%in Figure~\ref{fig:period-cycle}, that
%quiescent rates are apparently between those for Cycles 5 and 8.

%%%%%%%%%%%%%%%%%%%%%%%%%%%%%%%%%%%%%%%%%%%%%%%%%%%%%%%%%%%%%%%%%%%%%%
\subsection{\chandra\ HRC-I}
\label{sec:otherxray-hrci}
%%%%%%%%%%%%%%%%%%%%%%%%%%%%%%%%%%%%%%%%%%%%%%%%%%%%%%%%%%%%%%%%%%%%%%

The High Resolution Camera for Imaging (HRC-I)
is a microchannel plate detector with practically no
energy resolution but we include its two observations because:
they occurred three and a half years apart, near a maximum and minimum 
of our model 7-year cycle; \chandra\ is an active mission
and there may be more HRC-I observations for comparison in the future;
the first one overlaps with a \swift\ observation
(see Section~\ref{sec:otherxray-conversions}).
These two calibration observations
%, to study the telescope PSF off axis,
were piggybacked on primary observations
to measure the ACIS background while
ACIS was stowed out of the telescope light path.
% the HRC-I
%observations simultaneously collected data to study the off-axis PSF.
To do this, the instrument module was moved to a location where
the HRC-I could only observe Prox Cen far off the optical axis,
at 15.0' for ObsID 14276 and 25.62' for ObsID 17377.
This greatly broadened the source PSF, requiring large
elliptical extraction regions of 155\arcsec$\times$102\arcsec
and 354\arcsec$\times$216\arcsec, respectively.  
%14276: 315,207 pix = 155" x 102"
%17377: 720,440 pix = 354" x 216"
Roughly a quarter of the ObsID 17377
source-region counts during quiescence were from background,
less for 14276.
% $315\times207$ pixels for 14276 and $720\times440$ for 17377.  

During these observations the HRC-I operated in a limited telemetry mode
using only a portion of the detector.  The first 13 ks and last 1 ks
of the 50-ks Obs\-ID 17377 suffered telemetry saturation caused
by background `flares.'  The true source rates 
could not be accurately recovered during those times so they are excluded
from our analysis.  Telemetry was also saturated for roughly 1.5 ks
in the middle of the observation because of source flaring but
this does not affect our study of quiescent emission.
% 1/5 of normal dither (8" peak to peak).

Event rates are corrected for vignetting, which is a significant
effect so far off axis.  The correction factor is 0.753 at 15.0'
and taken from the \chandra\ calibration database.  The
CALDB vignetting tables only go to 20', but the vignetting factor
for 25.62' has been previously measured and modeled
to be 0.545.\footnote{
   %CXC memo `HRC-I and HRMA Calibration Using RX J1856.5-3754,'
   \url{http://cxc.harvard.edu/ccr/proceedings/02\_proc/presentations/bradw/rxj/}.
}
Lastly, because the HRC-I has no useful energy resolution, the rates plotted
in Figure~\ref{fig:ratedistrib}
refer to the full range of pulse heights rather than the
0.5--2.5 keV rates plotted for CCD instruments.

%%%%%%%%%%%%%%%%%%%%%%%%%%%%%%%%%%%%%%%%%%%%%%%%%%%%%%%%%%%%%%%%%%%%%%
\subsection{Determination of Quiescent Rates}
\label{sec:otherxray-quiet}
%%%%%%%%%%%%%%%%%%%%%%%%%%%%%%%%%%%%%%%%%%%%%%%%%%%%%%%%%%%%%%%%%%%%%%

\begin{figure}[t]
%\epsscale{0.50}
\epsscale{1.20}
\plotone{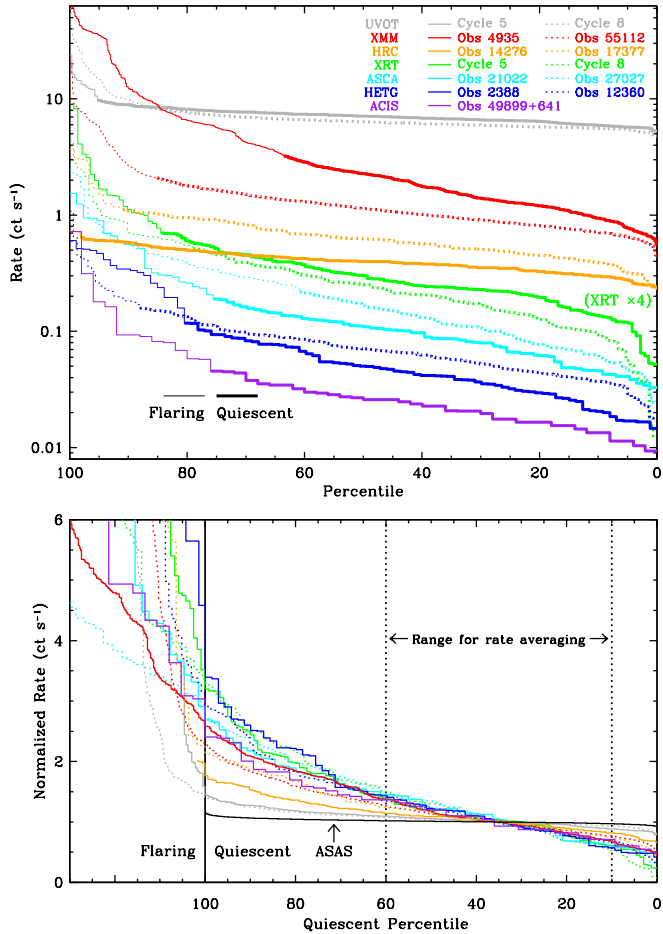}
\caption{
Light curve rate distributions.  (Top) Quiescent emission is
marked with thicker lines.  Rates have corrections for vignetting (HRC)
and deadtime (especially XMM),
but ACIS data are not corrected for exclusion of the heavily piled-up PSF core. 
XRT rates are multipled by four to avoid overlap with HETG data. 
(Bottom) Quiescent data are
rescaled along both axes: percentiles now refer only to quiescent emission,
and each observation's rates are normalized to the average rate in the
10--60\% quiescent range.  
ASAS optical $V$-band data are plotted for comparison,
and are treated as if they are all quiescent.
Note that instruments with less high-energy response 
(see Figure~\ref{fig:normarfs})
are less sensitive to short-term emission variability such as flares,
making it easier to ascertain the longer-term quiescent emission level.
%UVOT rates are not normalized with
%the X-ray data as their distribution is much flatter because
%of a relatively large baseline rate; 
%instead they have a constant offset applied to better illustrate 
%the significant difference in rates between \swift\ Cycles 5 and 8.
}
\label{fig:ratedistrib}
%\end{figure*}
\end{figure}

Rate distributions with instrumental adjustments
(enclosed energy fraction, livetime, vignetting) 
are plotted in Figure~\ref{fig:ratedistrib}.
% and  include corrections
% for instrument deadtime, which was generally $\sim$1\%
% except for XMM PN ObsIDs 4935 (30\% deadtime) and 55112 (5\%).
The top panel shows all the data and
illustrates the large variation in event rates among different instruments,
as well as the general shape of the rate distributions: relatively
flat for quiescent emission, and increasingly steep and unpredictable
for higher rate, less frequent flares.  Our interest here is
on the flattest part of the distributions, where rates are
relatively insensitive to the choice of sampling range.

Deciding where to draw the line between flaring and quiescent emission
is somewhat subjective, but normalizing the distributions along both
axes as shown in the bottom panel 
%(see caption for an explanation of the normalization methods) 
is helpful in guiding that judgment.
We iteratively adjusted
the flare/quiet break for each curve in the top panel and
plotted the results in the bottom panel, aiming to have the
curves overlap as much as possible, placing the highest emphasis on
the degree of overlap in the nearly linear 10-60\% quiescent range marked
with dotted lines (used to calculate our quiescent reference rates)
and least emphasis in the inherently more variable
flaring range.  

We were unable
to craft an automated method of doing this but believe our results
are reasonably objective.  We exclude the lowest 10\% 
from our calculations because of that
range's nonlinear rate distributions, which may be
caused in part by statistical artifacts from low-count binning, 
outlier source fluctuations, 
or instrumental/processing defects (particularly for
the XRT with its damaged CCD pixels).  
The upper limit of 60\% aims
to maximize the sampling basis 
while minimizing flare contamination.
% the probability of including a significant flare.
Changing the flare/quiet break by $\pm$10\% (using percentiles in the
top panel of Figure~\ref{fig:ratedistrib}) changes the reference
quiescent rates for X-ray instruments
% (not including the UVOT)
by typically 6\%, ranging from 10\% for \asca\ ObsID 27027
to 2\% for HRC-I ObsID 14276.
The corresponding UVW1 sensitivity is $\sim$1.3\%,
%As discussed later, 
%We believe that relative uncertainties in our reference rate
% Not sure what I meant here so remove::: These uncertainties
%are smaller than uncertainties arising from variations in
%Prox Cen's emission on time scales between a day and a year.
and the even flatter ASAS-3 distribution is shown for comparison.
%For comparison we also include a curve for the ASAS optical data.
%Apart from intensity changes, differences from year to year are 
%subtle and so the ASAS curve represents all nine years of data.

The sensitivity of inferred quiescent rates to the location
of the flare/quiet break is 
%easily understood in terms of
effectively given by 
the slope of the rate distributions in
the quiescent range (easiest to compare in the $\sim$50--100\% range
of the bottom panel of Figure~\ref{fig:ratedistrib}), 
e.g., the HRC-I distributions are the flattest of the X-ray data,
followed by \xmm\ and ACIS on up to the steepest distributions
of \asca\ and the HETG.
This is in turn highly correlated with 
the various instruments'
energy dependent effective areas (EAs), shown in Figure~\ref{fig:normarfs}.
Again using the HETG as an example,
its rate distributions have steep slopes 
while its EA is weighted heavily toward the higher
energies typical of emission from hot plasma, which shows more
intensity variation than emission at lower energies.  At the other extreme,
the HRC-I rate distributions are the flattest while the HRC EA is
more heavily weighted toward low energies where emission is less variable.
%recall from Figure~\ref{fig:crosscal} how flares have
%much less effect on rates in the HRC than the XRT.
% This figure comes later so it doesn't make sense to refer to it here.
All things otherwise being the same,
we would therefore expect HRC observations to be the most likely
to yield accurate quiescent rate measurements while instruments
with higher proportions of high-energy effective area, 
being more sensitive to high-$T$ flare emission, 
are less likely
to observed periods when high-$T$ (non-quiescent) emission is minor.

One weakness of the rate distribution analysis is that temporal
information is ignored.  One can see in Figure~\ref{fig:lcALL}
that the first two-thirds of \xmm\ ObsID 4935
very likely includes the slow decay of a large flare and so this
observation probably never sampled quiescent emission.
HETG ObsID 12360 exhibits a less obvious decline
but our estimated quiescent rate is again probably too high.  
The quiescent rate in \asca\ ObsID 27027
may also be overestimated.
%, and as noted before, \asca\ data
%have large calibration uncertainties at low energies.

%%%%%%%%%%%%%%%%%%%%%%%%%%%%
% FIGURE ##
% \label{fig:ratedistrib}
%%%%%%%%%%%%%%%%%%%%%%%%%%%%
% Use figure* for 2col
%%%%%%%%%%%%%%%%%%%%%%%%%%%%

%%%%%%%%%%%%%%%%%%%%%%%%%%%%%%%%%%%%%%%%%%%%%%%%%%%%%%%%%%%%%%%%%%%%%%
\subsection{Rate to Luminosity Conversions}
\label{sec:otherxray-conversions}
%%%%%%%%%%%%%%%%%%%%%%%%%%%%%%%%%%%%%%%%%%%%%%%%%%%%%%%%%%%%%%%%%%%%%%

With this analysis method we obtain UVW1 rates for \swift\ Cycles 5 and 8
of 6.618 and 6.061 ct s$^{-1}$ (a difference of 9.2\%) vs.\ 
% qsample.f does not correct for 1percent/year UVOT QE loss.
% I separately calculated that for Cycle 8 vs 5 and have applied
% a 2.85% correction above.
rates of 6.597 and 6.137 (difference of 7.5\%) 
obtained using the `quiescent snapshot' method
(see Table~\ref{table:swiftcycleaves}),
which is good agreement
given the various sources of uncertainty in both approaches.

\begin{figure}[t]
\epsscale{1.15}
\plotone{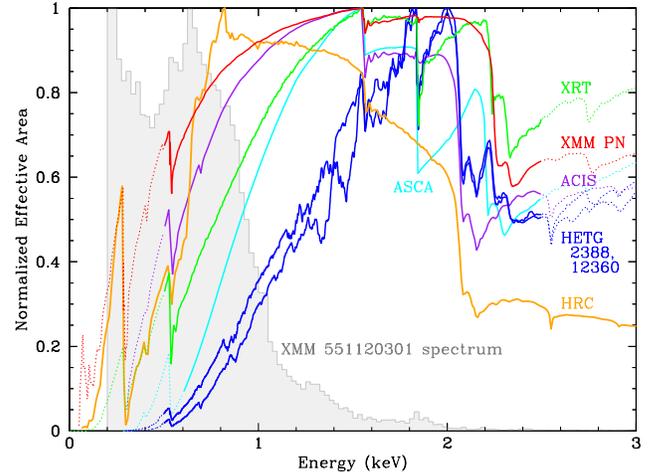}
\caption{
Normalized X-ray instrument effective areas, illustrating differences
in energy dependence.  Solid lines show the 0.5--2.5 keV energy range
used for rate measurements (0.6--2.5 keV for \asca); 
the HRC-I has no energy resolution so its
full range was used.
An \xmm\ spectrum is shown to
illustrate that most emission occurs at relatively low energies.
Instruments with more area at higher energies,
such as the HETG, detect relatively more high energy (more variable) 
emission.
}
\label{fig:normarfs}
%\end{figure*}
\end{figure}

\begin{figure}[t]
\epsscale{1.15}
\plotone{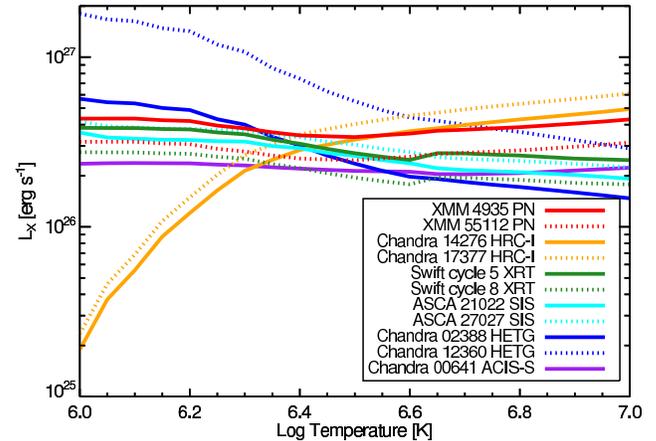}
\caption{
X-ray luminosities as a function of $T$, derived from measured
quiescent reference rates via PIMMS.
}
\label{fig:LUMvsT}
%\end{figure*}
\end{figure}

To compare emission observed by the many X-ray instruments
we must convert their average
quiescent event rates to 0.5--2.5 keV luminosities, which we did using the
Portable Interactive Multi-Mission Simulator (PIMMS; v.~4.8)
tool.\footnote{
  \url{https://heasarc.gsfc.nasa.gov/docs/software/tools/pimms.html}
  } 
As noted before, instrument responses can vary a great
deal as a function of energy
(see Figure~\ref{fig:normarfs}) and, to a lesser degree, time
(for \chandra\ and \asca).
%Figure LUMvsT isn't ready--drop it.
PIMMS-derived luminosities are plotted in Figure~\ref{fig:LUMvsT},
showing that results for some instruments, 
particularly the HRC-I and HETG, depend strongly on the assumed temperature.
For consistency, and because some spectra did not have enough counts
to permit detailed modeling, we used a single set of plasma
parameters for all the PIMMS rate conversions.  Those values were derived
from fits to \xmm\ spectra, which had by far the most counts.
Auxiliary Response Functions (ARFs) and 
Response Matrix Functions (RMFs) were created for ObsIDs 4935 and 55112
using the Scientific Analysis System v.~1.2 (SAS)\footnote{
        \url{http://www.cosmos.esa.int/web/xmm-newton/sas}
        }
{\tt evselect}
command, with standard {\tt flag=0} and {\tt pattern=0:4} (sd) event filtering.

%%%%%%%%%%%%%%%%%%%%%%%%%%%%
% FIGURE ##
% \label{fig:normarfs}
%%%%%%%%%%%%%%%%%%%%%%%%%%%%
% Use figure* for 2col
%%%%%%%%%%%%%%%%%%%%%%%%%%%%

%%%%%%%%%%%%%%%%%%%%%%%%%%%%
% FIGURE ##
% \label{fig:LUMvsT}
%%%%%%%%%%%%%%%%%%%%%%%%%%%%
%% Use figure* for 2col
%%%%%%%%%%%%%%%%%%%%%%%%%%%%%
        
To fit the spectra we used the Sherpa modeling and fitting package
\citep{cit:freeman2001} with a two-temperature APEC coronal emission model.
Column density was set to $10^{18}$ cm$^{2}$, providing negligible
absorption.
The best fit to the ObsID 5512 quiescent spectrum was obtained using
70\% $kT_1=0.23$ keV, 30\% $kT_2=0.80$ keV, and 0.25 solar abundance.
Results for ObsID 4935 were similar but with higher
flux.  Because of previously noted concerns over whether
that observation includes truly quiescent emission used the
ObsID 55112 results in all our PIMMS calculations.

%As can be seen, the rate-to-luminosity conversions are roughly
%independent of $T$ for the CCD measurements, but have some $T$
%dependence for the HETG observations because of the additional
%wavelength dependence of the gratings, and even more dependence
%for the HRC observations.  The latter effect is because the HRC
%has relatively more of its effective area at low energies than
%the CCD instruments.  

\begin{figure}[t]
\epsscale{1.15}
\plotone{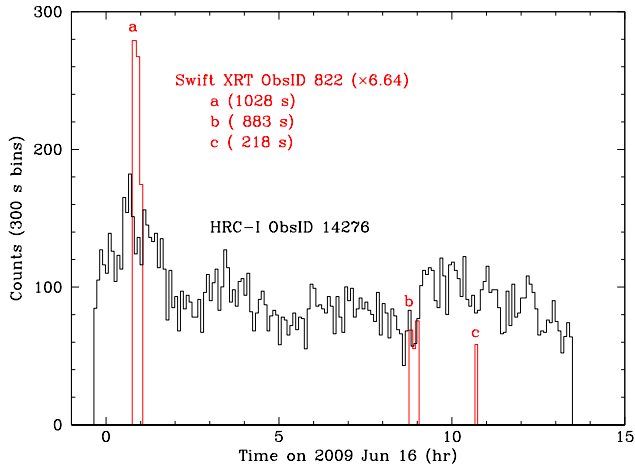}
\caption{
	Background-subtracted light curves of overlapping 
	\swift\ XRT and \chandra\ HRC observations.  
%	300-s bins were used for the HRC, while each
%	\swift\ snapshot was evenly divided into an integer number of bins,
%	with rates scaled to their 300-s equivalents.
	\swift\ snapshots were each evenly divided into an integer number
	of bins and their counts (0.5--2.5 keV) rescaled for 300-s bins.
	Events during \swift\ snapshots 822b and 822c
	were used to cross calibrate the two instruments.
%	, yielding 
%	a ratio of event rates $R_{HRC}/R_{XRT}=6.64\pm1.33$ (0.5--2.5 keV
%	for XRT, full range for HRC) for quiescent emission.  
	The rate difference during snapshot 822a
	is because the XRT's CCD detector is more sensitive than the
	HRC to the higher-energy emission from hotter (flaring) plasma
	(see Figure~\ref{fig:normarfs}).
}
\label{fig:crosscal}
%\end{figure*}
\end{figure}

Even with what should be well determined plasma parameters,
the sensitivity of the HRC-I rate-to-luminosity calculations 
to the assumed temperature is
a concern, especially in light of the challenges faced by
\citet{cit:ayres2009} when comparing measurements from
different instruments (see Section~\ref{sec:intro-xrays}). 
Luckily, one of the \swift\ observations overlaps with the
first HRC observation, as seen in Figure~\ref{fig:crosscal}.
Two of the three ObsID 822 snapshots collected quiescent emission,
allowing a direct cross calibration of HRC and XRT rates.
During the time of overlap, the HRC collected 333 events
(with estimated 95 background) versus 39 (3 background) for the XRT
in the 0.5--2.5 keV range,
yielding a ratio of $6.64\pm1.33$.
%This is only half the ratio derived from the PIMMS calculations,
%indicating that ...***something... related  to T?
Figure~\ref{fig:crosscal} 
also nicely illustrates that the HRC is significantly
less sensitive to emission from flares than the \swift\ XRT 
and other instruments that have more of their effective area 
at higher energies than the HRC (see Figure~\ref{fig:normarfs}).
Note that using the PIMMS-derived HRC-I curves in
Figure~\ref{fig:LUMvsT} would yield luminosities roughly
twice the values we compute from cross-calibration with the XRT;
we have no obvious explanation, but again point to the
difficulties of reconciling results from instruments
with different energy responses.

%recall from Figure~\ref{fig:crosscal} how flares have
%much less effect on rates in the HRC than the XRT.

%%%%%%%%%%%%%%%%%%%%%%%%%%%%
% FIGURE ##
% \label{fig:crosscal}
%%%%%%%%%%%%%%%%%%%%%%%%%%%%
% Use figure* for 2col
%%%%%%%%%%%%%%%%%%%%%%%%%%%%

%%%%%%%%%%%%%%%%%%%%%%%%%%%%%%%%%%%%%%%%%%%%%%%%%%%%%%%%%%%%%%%%%%%%%%
\subsection{Results and Uncertainties}
\label{sec:otherxray-results}
%%%%%%%%%%%%%%%%%%%%%%%%%%%%%%%%%%%%%%%%%%%%%%%%%%%%%%%%%%%%%%%%%%%%%%

\begin{figure}[t]
\epsscale{1.15}
\plotone{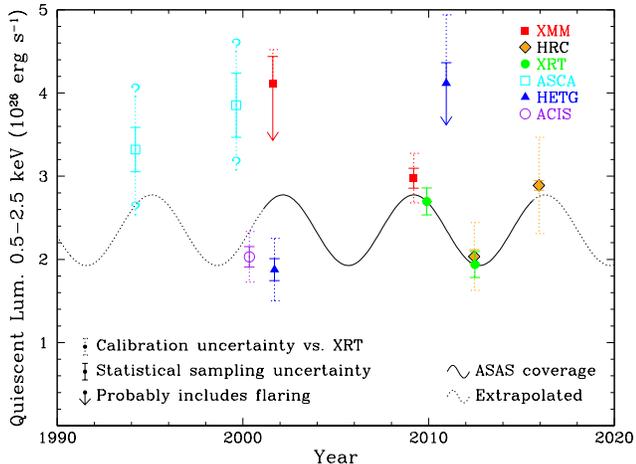}
\caption{
Quiescent X-ray luminosities (0.5--2.5 keV) over time, 
with the optical 7.05-yr cycle scaled to match XRT data.
See text and Table~\ref{table:lumuncerts} for details
regarding error bars.
}
\label{fig:xfluxes}
%\end{figure*}
\end{figure}

The resulting quiescent X-ray luminosities measured over
the past 22 years are shown in Figure~\ref{fig:xfluxes},
with the optical 7-yr cycle scaled to intercept the
two \swift\ XRT points.
Error bars reflect information in Table~\ref{table:lumuncerts}.
For each point, solid error bars denote `statistical sampling' uncertainties
arising from the choice of the flare/quiet break in
Figure~\ref{fig:ratedistrib} (see Section~\ref{sec:otherxray-quiet}).
Dotted error bars are estimated uncertainties from cross-calibration
with the \swift\ XRT.  Based on work by the 
International Astronomical Consortium for High Energy Calibration 
\citep[IACHEC; e.g.,][]{cit:tsujimoto2011,cit:plucinsky2016},
these latter errors are typically $\sim$10\% at energies
greater than about 0.9 keV (15-20\% around 0.6 keV)
for missions launched after the mid 1990's, but 
the relatively high sensitivity of the \chandra\ HETG observations to
temperature uncertainties 
%(see Figure~\ref{fig:LUMvsT})
and the special treatment required for the piled-up ACIS observations
lead us to increase 
uncertainties for these instruments.

%%%%%%%%%%%%%%%%%%%%%%%%%
\begin{deluxetable}{cccc}
\tabletypesize{\scriptsize}
\tablecaption{Luminosity uncertainties
        \label{table:lumuncerts}}
\tablewidth{0pt}
\tablehead{\colhead{Observation}
                & \colhead{Calibration}
                        & \colhead{Statistical}
                                & \colhead{Subjective}	\\
                & \colhead{error}
                        & \colhead{sampling}
                                & \colhead{sampling}	\\
                & \colhead{vs.~XRT}
                        & \colhead{error}
                                & \colhead{error}	\\
                & \colhead{(\%)}
			& \colhead{(\%)}
                        	& 			
}
\startdata
\asca\ 21022		& ?	& 8	& low		\\
\asca\ 27027		& ?	& 10	& medium	\\
ACIS 49899$+$641 	&15	& 6	& low		\\
HETG 2388		& 20	& 7	& low		\\
HETG 12360		& 20	& 6	& high		\\
\xmm\ 4935		& 10	& 8	& very high	\\
\xmm\ 55112		& 10	& 4	& low		\\
\swift\ XRT Cycle 5	& ...	& 6	& very low	\\
\swift\ XRT Cycle 8	& ...	& 8	& very low	\\
HRC-I 14276		& 20	& 4	& low		\\
HRC-I 17377		& 20	& 2	& low		\\
\enddata
\tablecomments{
        Statistical sampling error is the change in the calculated quiescent
	rate when the flare/quiet break in Figure~\ref{fig:ratedistrib}
	changes by 10\%.  Subjective sampling error reflects the
	likelihood that the presumed quiescent emission includes significant
	contamination from flares.
        }
\end{deluxetable}
%%%%%%%%%%%%%%%%%%%%%%%%%

As noted earlier, \asca's calibration uncertainties are rather
large at low energies and increased over time.
We include its measurements 
%(divided by 2.5 so they fit on the plot)
in Figure~\ref{fig:xfluxes} but they should be given little weight.
HRC-I luminosities are even more sensitive to $T$ uncertainties
than the HETG, but our direct calibration vs.\ the XRT is accurate to 20\%.
Relative calibration uncertainty between the two HRC-I observations
is tiny because the effective area is very nearly constant, and
Prox Cen's quiescent luminosity during ObsID 17377 (2015 Dec)
is clearly higher than during ObsID 14276 (2012 Sep).

%%%%%%%%%%%%%%%%%%%%%%%%%%%%
% FIGURE ##
% \label{fig:xfluxes}
%%%%%%%%%%%%%%%%%%%%%%%%%%%%
% Use figure* for 2col
%%%%%%%%%%%%%%%%%%%%%%%%%%%%

Subjective sampling errors are assigned based on judgements of
the likelihood that our quiescent rates
may be incorrect (generally meaning too high because
of the inclusion of flare emission).  Measurements are most
reliable when they come from long,
multiple observations over a period of time.  In both cases,
the key advantage is a higher probability of observing emission during
periods of quiescence.  Multiple observations, such as \swift's,
that span periods comparable to or longer than Prox Cen's 
83-d rotation period
have the additional advantage of sampling emission over more of the
stellar surface.
In practice, given Prox Cen's propensity for flaring 
and the spatial nonuniformity that gives rise to its 
rotational intensity modulation, 
there will always be some ambiguity 
in what constitutes `quiescent' emission, but the \swift\
observations should provide the best measurements and we
assign them a `very low' sampling error in Table~\ref{table:lumuncerts}.

At the other extreme,
as noted at the end of Section~\ref{sec:otherxray-quiet},
\xmm\ ObsID 4935 probably had significant flare contamination
during its entire exposure, and our quiescent rates estimates for
HETG ObsID 12360
% and to a lesser extent \asca\ ObsID 27027 
are also likely to be too high.
%All three of these measurements are marked in 
Both of these measurements are marked in 
Figure~\ref{fig:xfluxes} with upper limits.

For these reasons we assign the most significance to the \swift\ XRT data,
followed by the HRC measurements, all of which are
in good (anti-correlated) accord with the 7-yr optical cycle,
as are the \swift\ UVOT/UVW1 measurements.
Observations by other X-ray missions are, after considering the
likelihood of flare contamination in some measurements, also consistent with
a cycle, although given the estimated uncertainties one cannot
draw too many conclusions.  There are also uncertainties from
extrapolation of the optical cycle to times before the first
ASAS data in late 2000; as illustrated particularly well by the
Sun's most recent cycles, there can be significant differences in period
and amplitude from one cycle to the next.

%%%%%%%%%%%%%%%%%%%%%%%%%%%%%%%%%%
\section{SUMMARY}
\label{sec:conclusions}

We have presented an analysis of 15 years of 
ASAS $V$-band optical monitoring data on Proxima Cen,
finding strong evidence for
%by \citet{cit:jason2007} and \citet{cit:guinan2010}
periodic 7-yr variations
and confirming previous measurements of an 83-d rotation period
by \citet{cit:benedict1998}, \citet{cit:kiraga2007},
\cite{cit:savanov2012}, and \cite{cit:suarez2016}.
We do not see any evidence for the 1.2-yr or 3-yr periodicities tentatively
reported by \citet{cit:cincu2007} or \citet{cit:benedict1998},
respectively, but our 7.05-yr optical period is in accord with
the intriguing peak around 7 yrs noted by \citet{cit:endl2008}
in their analysis of radial velocity data,
and with the $6.8\pm0.3$ yrs derived by \cite{cit:suarez2016}
%suggestions of periodicity
%by \citet{cit:jason2007} and \citet{cit:guinan2010}
%based on smaller sets of ASAS data.
from a smaller set of ASAS data.

The amplitude of $V$-band rotational modulation 
was observed to vary significantly on few-year time scales 
but the phase of the variations was remarkably consistent.
The lack of $I$ band variation combined with a strong trend in $V-I$ vs.~$V$,
with the star growing redder when fainter, imply that Prox Cen 
likely has a significant filling factor of cool starspots.
ASAS $V$-band data show evidence for differential rotation, in the
form of distinct $P_{\rm rot}$ values in different epochs.  The
fractional DR rate is $\Delta P_{\rm rot}/\langle P_{\rm rot} \rangle
\sim 0.16$, similar to the solar value and broadly consistent with
observed trends in single dwarfs \citep{cit:saar2011}

Our analysis of four years of \swift\ data (2009--2013) strengthens
the case for a stellar cycle by extending it to higher energies,
with observed peak-to-peak variations of order 10\% in the UVW1 band
and roughly a factor of 1.5 in the 0.5--2 keV X-ray band,
with X-ray/UV variations anti-correlated with optical brightness.
This anti-correlation is also seen (with less confidence)
in rotational modulation,
as would be expected if higher starspot coverage (which generates
more X-ray/UV emission) causes a net decrease in optical emission.
Comparing against six other stars with measured X-ray cycles,
we find that cycle amplitude correlates with Rossby number
according to $L_{\rm X}^{\rm max}/L_{\rm X}^{\rm min} \propto Ro^{1.39}$,
indicating that the X-ray cycle amplitude decreases with increasing
coronal activity, consistent with the idea that higher activity
stars have a greater fraction of their surfaces covered by
active regions and therefore less potential to increase
X-ray emission at cycle maxima.

Two recent \chandra\ HRC-I observations, one of which
occurred during a \swift\ observation allowing accurate
cross calibration, extend X-ray coverage to late 2015 and
are in excellent accord with the presumed cycle,
% based on \swift\ results,
as is the most recent \xmm\ measurement in 2009.
Our most reliable measurements therefore
%, by the HRC and \swift,
now cover two cycle maxima and one minimum.
%but more observations, ideally by \swift\ which is particularly
%well suited for this type of monitoring program, are needed
%in order to make a definitive claim of an X-ray cycle.
Other data from previous and currently operating X-ray missions
extending back more than two decades yield more ambiguous results,
illustrating the difficulty of measuring quiescent emission in
active stars such as Prox Cen
when observations are few and infrequent.
%from which we conclude that multiple observations per year are
%highly desirable in order to accurately determine the underlying quiescent
%emission level in the absence of flares, rotational modulation,
%and other short-to-moderate term variability in active stars such as Prox Cen.
Complications when comparing results from different instruments
were also highlighted.

%Although the ASAS and \swift\ data barely overlap, prohibiting
%direct analysis of intensity correlations, we find that sinusoidal
%fits to rotationally phased data
%%data binned according to the rotational period of the ASAS data 
%show that optical brightness variations are $\sim$exactly out of
%phase with X-ray and UV variations, 
%just as for the 7-yr cyclical variations.

The apparent 7-yr stellar cycle in Prox Cen, a fully convective
M5.5 star, is in conflict with
most models of magnetic dynamo theory and should spur further
theoretical work in this area.  
%The higher degree of optical variation
%and lower degree of X-ray variation over the cycle compared to
%the Sun may be important clues to guide those models.
Further evidence that dynamo behavior in fully-convective stars does 
not follow canonical theory is provided by
\citet{cit:wright2016},
who found that the X-ray emission of four four fully convective stars, 
including Prox Cen, correlates with Rossby number in the same way 
as in solar-type stars.  The X-ray activity-rotation relationship has 
long been established as a proxy for magnetic dynamo action;
these results, combined with our finding of Proxima's stellar cycle,
therefore suggest that fully convective stars operate dynamos
similar to that of the Sun, with the implication 
that a radiative core and its tachocline are not critical or 
necessary ingredients.

Our study of fifteen years and 1085 nights of ASAS $V$-band optical photometry,
three years of I-band observations, 
four years and 125 \swift\ X-ray and UV exposures,
and two decades of observations by other X-ray missions
comprises by far the most extensive analysis of long-term monitoring data
on an M dwarf and also provides the best evidence for a stellar cycle
in an isolated fully convective star.  
The ASAS-4 monitoring program is continuing to collect data and the
All-Sky Automated Survey for Supernovae 
\citep[ASAS-SN;][]{cit:shappee2014}
obtained its first observation of the field containing Prox Cen
on 2016 Mar 9 (private communication, B.~Shappee), so there
are excellent prospects for sustained optical monitoring of this star.

Additional X-ray data would be even more valuable but are 
harder to obtain than optical data, and determinations of quiescent
luminosities are a challenge because of frequent flaring.
Our work shows that reliable measurements of quiescent emission
can be made even when monitoring active stars such as Prox Cen,
but that this is most easily accomplished when there are
several observations per year that sample all sides of the star, 
made by the same instrument
(or multiple instruments with good cross calibration), and preferably 
in softer energy ranges less sensitive to flares.
Each observation can be quite short, however, so that
with the proper instrument(s), a modest investment of
observing time can yield UV and X-ray data
vital for the study of cyclic and other medium- to long-term stellar behavior.

%Short-term XUV
%variability from flaring makes looking for longer-term
%trends a challenge, but careful construction of observing programs
%and analysis methods can yield measurements that lead to important insights
%that can not be obtained in any other way.

%%%%%%%%%%%%%%%%%%%%%%%%%%%%%%%%%%%%%%%%%%%%%%%%%%%%%%%%%%%%%%%%%%%%
 
\acknowledgments

This work was supported by NASA's \swift\ Guest Investigator
program under Grants NNX09AR09G and NNX13AC61G.
BJW, JJD, and VLK were also supported by NASA contract NAS8-39073 to
the \chandra\ X-Ray Center,
and SHS was supported by NASA Heliophysics grant NNX16AB79G.
We thank the \swift\ team and especially the PI, Neil Gehrels,
for providing TOO/Discretionary time, without which much of
this work would have been impossible.
We also thank Ben Shappee for helpful conversations
and the ASAS collaboration for providing optical photometry data.
This work made use of data supplied by the UK \swift\ Science Data 
Centre at the University of Leicester, and
data and software provided by the 
High Energy Astrophysics Science Archive Research Center (HEASARC), 
which is a service of the Astrophysics Science Division at NASA/GSFC 
and the High Energy Astrophysics Division of the 
Smithsonian Astrophysical Observatory.

%%%%%%%%%%%%%%%%%%%%%%%%%%%%%%%%%%%%%%%%%%%%%%%%%%%%%%
%  REFERENCES
%%%%%%%%%%%%%%%%%%%%%%%%%%%%%%%%%%%%%%%%%%%%%%%%%%%%%%
%\newpage
% \section{References}
% \label{sec:references}

% now the references. delete or change fake bibitem. delete next three
% lines and directly read in your .bbl file if you use bibtex.
% \begin{references}

% Comment out the next line before submitting
% \small

%%%%%%%%%%%%%%%%%%%%%%%%%%%%%%%%%%%%%%%%%%%%%%%%%%%%%%

%%%%%%%%%%%%%%%%%%%%%%%%%%%%%%%%%%%%%%%%%%%%%%%%%%%%%%
\end{document}